\documentclass[11pt, a4paper, twocolumn, twoside]{article}
\usepackage{fancyhdr}
\usepackage{graphicx}
\usepackage{booktabs}
\usepackage{multirow}
\usepackage{moresize}
\usepackage{subfig}
\usepackage{amsmath}
\usepackage{amssymb}
\usepackage{amsfonts}
\usepackage{abstract}

\title{A Comparison of new MC-adapted Parton Densities}
\author{T.~Kasemets\thanks{tomas@kasemets.se} \hspace{1mm}and T.~Sj\"{o}strand\thanks{torbjorn@thep.lu.se}  \\\\
\textit{Theoretical High Energy Physics}\\
 \textit{Department of Astronomy and
Theoretical Physics,}\\ \textit{Lund University,} \\\textit{S\"olvegatan 14A,}\\ \textit{SE 223 62 Lund, Sweden}\\[0.3cm]}
\date{}

\fancypagestyle{plain}{
\fancyhf{}
\fancyfoot{}
}
\advance\textwidth +40pt
\advance\oddsidemargin -20pt
\advance\evensidemargin -20pt
\advance\topmargin -25pt
\advance\textheight +50pt
\newcommand{\stext}[1]{\mbox{\ssmall{#1}}}

\begin{document}
\pagestyle{fancy}
\twocolumn[

\begin{flushright}
LU TP 10-19\\
MCnet/10/13\\
July 2010
\end{flushright}
\maketitle

\begin{onecolabstract}
A selection of the latest and most frequently used parton distribution
functions (PDFs) is incorporated in
\textsc{Pythia8}, including the Monte Carlo-adapted PDFs from the MSTW and CTEQ collaborations. This article examines the differences
in PDFs as well as the effect they have on results of simulations and compare with data
collected by the CDF experiment. Monte Carlo-adapted PDFs do a better job than leading- and next-to-leading order PDFs for many observables, but there is room for further improvements.\\
    \end{onecolabstract}
  ]
\saythanks
\section{Introduction}

Precision measurements of cross sections should be compared
with precision theoretical calculations. The state of the art
is next-to-leading order (NLO) calculations of matrix elements
(MEs), which then should be combined with NLO parton distribution
functions (PDFs) to obtain the theoretical predictions for
various processes at hadron colliders. Such an approach works well
for sufficiently inclusive quantities, say the total cross
section for $W^{\pm}$ production.

At the other extreme, say for the production of jets associated
with the $W^{\pm}$, experimental jet finding will be based on
the clustering of a complex hadronic final state. Currently
such states can only be modeled by the use of event generators,
where many components are only formulated to leading order (LO),
such as multiparton interactions (MPIs), initial-state radiation
(ISR) and final-state radiation (FSR). In these descriptions it is thus not apropriate to use NLO PDFs.

To be more specific, the combination of LO MEs with NLO PDFs is
accurate to LO, just as LO MEs with LO PDFs would be. So from
a formal point of view the use of NLO or LO PDFs is equivalent.
There is a key point, however: LO PDFs have a simple
probabilistic interpretation and are always positive,
as are LO MEs. At NLO positivity is no longer required, neither
for PDFs nor for MEs. The convolution of the two hopefully should
still be positive, but even that is strictly not guaranteed over
all of phase space. Characteristic for NLO PDFs is especially
that the gluon distribution at small $x$ and $Q^2$ tends to come
out negative, or at least very small, in order to describe the evolution of
$F_2$ with $Q^2$.
This means that the LO ME $+$ NLO PDF combination breaks down
in this region, which is where most of the underlying-event activity
originates (from MPIs, ISR and FSR).

Thus there is a need to continue the development and use of LO PDFs.
In recent years this field has received renewed attention \cite{mrstmod}-\cite{cteqmc}.
In particular, attempts have been made to find LO PDFs that,
when combined with LO MEs, attach better to the complete NLO
behavior for a selection of cross sections. This is accomplished primarily by
relaxing the momentum sum rule, in order to allow the PDFs to have a large
small-$x$ gluon distribution without compromising the large-$x$ quark
distributions. Studying these new
PDFs can shed more light on which features of the PDFs are ideal for
leading-order MC generators and how further improvements can be made. 

To this end we have incorporated ten new PDFs into \textsc{Pythia8} \cite{Pythia8} and compare the
new MC-adapted PDFs with regular LO and NLO PDFs. We also study how the differences between them affect results of
simulations, and compare the results with data collected by the CDF
experiment. The observables studied are somewhat different from the ones used
in the construction of the PDFs and thus give a complementary picture.

The
structure of the article is as follows. Section \ref{sec:PDFPYTHIA} describes the inclusion of
the PDFs in \textsc{Pythia8}. The different PDFs are compared in section \ref{sec:Compare}. 
The results from simulations of minimum-bias events and hard QCD
events are studied in sections \ref{sec:minbias} and \ref{sec:jet}, respectively. Finally the article
concludes with a summary and outlook in section \ref{sec:sum}.

\section{PDFs in PYTHIA8}
\label{sec:PDFPYTHIA}
\textsc{Pythia8} has so far been distributed with the option to choose between two PDFs,
GRV94L \cite{grv} and CTEQ5L \cite{cteq5l}, which are both fairly old. Many
new and improved PDFs have been released and made available to \textsc{Pythia8}
simulations only through LHAPDF \cite{lhapdf}. The LHAPDF package has grown quite large and in the process also somewhat slow, and the code is written in Fortran while the community is changing to C++. It is desirable to include some PDFs directly into \textsc{Pythia8} since it can speed up simulations, make \textsc{Pythia8} more complete and facilitate the switching between different frequently used PDFs. We therefore incorporate
ten new PDFs from the MRST \cite{mrstmod}-\cite{mrstmod2}, MSTW 2008 \cite{mstw2008}, CTEQ6
\cite{cteq6} and CTEQ MC \cite{cteqmc} distributions into \textsc{Pythia8},
listed in Tab.~\ref{pdfrange}. Two of them
are NLO, which are not intended for LO MC use, but included for
comparison. Inclusion of the PDFs was done in
close contact with the MSTW and CTEQ collaborations.

Including additional PDFs proved to be less straightforward than might first
be expected, see further below. A major reason for this is the need to, in MC simulations, go
outside the range of the PDF grids; specifically to
smaller $x$ and $Q^2$ values. MSTW provides routines not only for interpolation but also
for extrapolation outside this grid, while the CTEQ collaboration has
recommended a freeze of the PDFs at the value just inside the
grid. The range of the grids for the different PDFs are shown in
Tab.~\ref{pdfrange}, together with their $\alpha_S$ values and running.

The code supplied by the authors had to be modified to fit natively
into \textsc{Pythia8} and we carried out extensive tests. When possible the tests included
comparisons with the corresponding PDFs in the LHAPDF package. The
\textsc{Pythia8}-included PDFs run about a factor two faster than they do going the
way via the LHAPDF package.
 
\begin{table*}
  \center
  \begin{tabular}{lllll}
    \toprule
    PDF & $x$ range & $Q^2$ range [GeV$^2$] & $\alpha_S$ & $\alpha_S(M_Z)$
    \\
    \midrule
    GRV94L & $10^{-5} - 1$ & $0.40 - 10^6$ & LO & 0.128 \\ 
    CTEQ5L & $10^{-6} - 1$ & $1.00 - 10^8$ & LO & 0.127 \\
    MRST LO* & $10^{-6} - 1$ & $1.00 - 10^9$ & NLO & 0.12032 \\
    MRST LO** & $10^{-6} - 1$ & $1.00 - 10^9$ & NLO & 0.11517 \\
    MSTW LO & $10^{-6} - 1$ & $1.00 - 10^9$ & LO & 0.13939 \\
    MSTW NLO & $10^{-6} - 1$ & $1.00 - 10^9$ & NLO & 0.12018 \\
    CTEQ6L & $10^{-6} - 1$ & $1.69 - 10^8$ & NLO & 0.1180 \\
    CTEQ6L1 & $10^{-6} - 1$ & $1.69 - 10^8$ & LO & 0.1298 \\
    CTEQ66 (NLO)  & $10^{-8} - 1$ & $1.69 - 10^{10}$ & NLO & 0.1180 \\
    CT09MC1 & $10^{-8} - 1$ & $1.69 - 10^{10}$ & LO & 0.1300 \\
    CT09MC2  & $10^{-8} - 1$ & $1.69 - 10^{10}$ & NLO & 0.1180 \\
    CT09MCS  & $10^{-8} - 1$ & $1.69 - 10^{10}$ & NLO & 0.1180 \\
    \bottomrule
  \end{tabular}
  \caption{The PDFs now included in \textsc{Pythia8}, with the $x$ and $Q^2$
    ranges of the respective grids, as well as the order of the running of 
    $\alpha_S$ and the value at $M_Z$}
  \label{pdfrange}
\end{table*}

\subsection{MRST/MSTW}
The PDFs supplied to us
from MSTW were in some respects improved compared to the versions available
in LHAPDF. Our implementation for the
MRST LO* and LO** PDFs makex use of the new MSTW grid ($64\times48$) ranging
down to $x=10^{-6}$ while the LHAPDF
versions use the original grid with fewer ($49\times37$) grid points and
shorter $x$ range ($x_{min}=10^{-5}$). The new grid results in a less steep gluon distribution towards small $x$, and at $x=10^{-8}$ the difference reaches a factor of two. The values of $\alpha_S$ are slightly different in the new LO* and LO** grid files
than in the corresponding LHAPDF ones. The LHAPDF versions use
$\Lambda_{QCD}$ for four active flavors, and the change to $\Lambda_{QCD}$ for five active flavors yields a slightly different
value. Also worth noticing is that LO* and LO** both use
the unorthodox value of the $Z$ boson mass, $M_Z = 91.71$~GeV, unlike the MSTW 2008 distribution which uses $M_Z=91.19$~GeV \cite{lhapdf}.
For MSTW 2008 LO, LO* and LO** the interpolation gave negative gluon values
 at some large-$x$ intervals and for a wide range of $Q^2$ scales. The gluon distribution is very small in this region and therefore
the negative values do not affect the results of the simulations. Furthermore, LHAPDF can give
negative values for the up quark for $0.9 < x < 1.0$ at large $Q^2$, which is worse since the up quark dominates
for large $x$ values. 

The MSTW NLO distribution gave very large negative values for the anomalous dimension $\frac{d\log(xf)}{d\log(Q^2)}$ at small
$Q^2$ and $x$ values around $10^{-5}$, which resulted in a huge
$\bar{s}$ distribution when extrapolated to low $Q^2$. This could also be a problem for the gluon which could
get large negative anomalous dimensions in
the $x$ region where the distribution is negative. To avoid this the anomalous
dimension is manually forced to be larger than $-2.5$. Although this does fix
the issue at hand, it is also an example of the dangers of using NLO PDFs in LO MC simulations, and an indication that
one has to be very careful with such use.
\subsection{CTEQ 6/MC}
The CTEQ distributions work well inside the grid but outside or near
the edges some problems occurred. The $tv = \log(\log(Q))$ values in the grid file were discovered not to exactly
correspond to the $Q$ values in the same file, and hence some points inside the $Q$
grid would end up outside the $tv$ grid. This caused some strange errors, for example
the $b$ quark distribution, after being zero below the threshold, suddenly became
huge at $Q^2$ values just inside the grid. Therefore we choose to read in only the
$Q$  grid points and then calculate $tv$.

Outside the grid there can be differences between the CTEQ6 PDFs in \textsc{Pythia8} and the corresponding
ones in LHAPDF. This is because LHAPDF provides the option to
use extrapolation
routines where \textsc{Pythia8}, by recommendation from the CTEQ authors, freezes the
values.

\section{Comparison of PDFs}
\label{sec:Compare}
There are strong similarities between the PDFs but also large differences, especially in the small $x$ region. Broadly the LO PDFs are similar, except for MSTW LO which is much larger than the others for small $x$ values, the MC-adapted PDFs are similar and the NLO PDFs are similar. Comparing the two groups, the CTEQ distributions have smaller distributions at small $x$, both for the PDFs that freeze and for the ones with the grid ranging down to $10^{-8}$, except for the NLO distributions. The two NLO PDFs have a smaller gluon distribution at small $x$, and MSTW NLO is negative at $Q^2=4$~GeV$^2$.

The gluon distribution is dominating in the region of small $x$ while the
valence quarks, and then especially the up quarks, dominate for large $x$. For
a first comparison we
therefore choose to focus mainly on these two distributions.

The PDFs are different from one another in several aspects.
Four of them do not obey the momentum sum rule.
LO* carries 1.12 times the proton momentum, LO** a $Q^2$-dependent
number between 1.17 and 1.14, MC1 1.10 and MC2 1.15. The special
behavior of LO** is related to the use of $\alpha_S(p_{\perp}^2)$
instead of $\alpha_S(Q^2)$ in the evolution. The MC-adapted distributions
from CTEQ have pseudodata from full NLO calculations included in their
fit. MCS is the only MC-adapted PDF that does not break the momentum sum rule,
but instead has
more freedom in the parameterization: renormalization and factorization
scales are allowed to vary in the global fit.

\begin{figure*}[tp]
  \centering
  \subfloat[]{\label{fig:uxQ24-1}\includegraphics[width=0.5\textwidth]
    {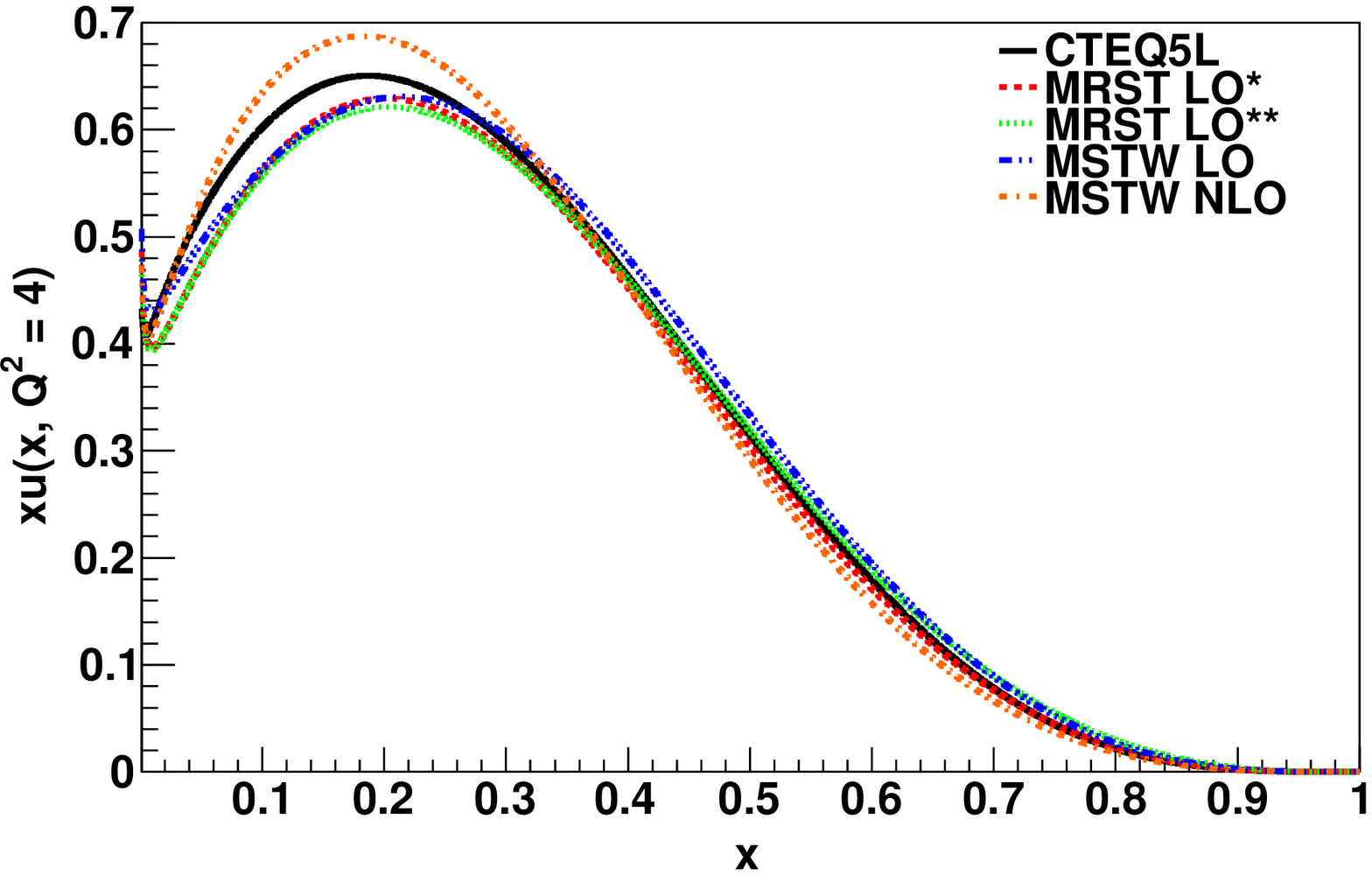}}                
  \subfloat[]{\label{fig:uxQ24-2}\includegraphics[width=0.5\textwidth]
    {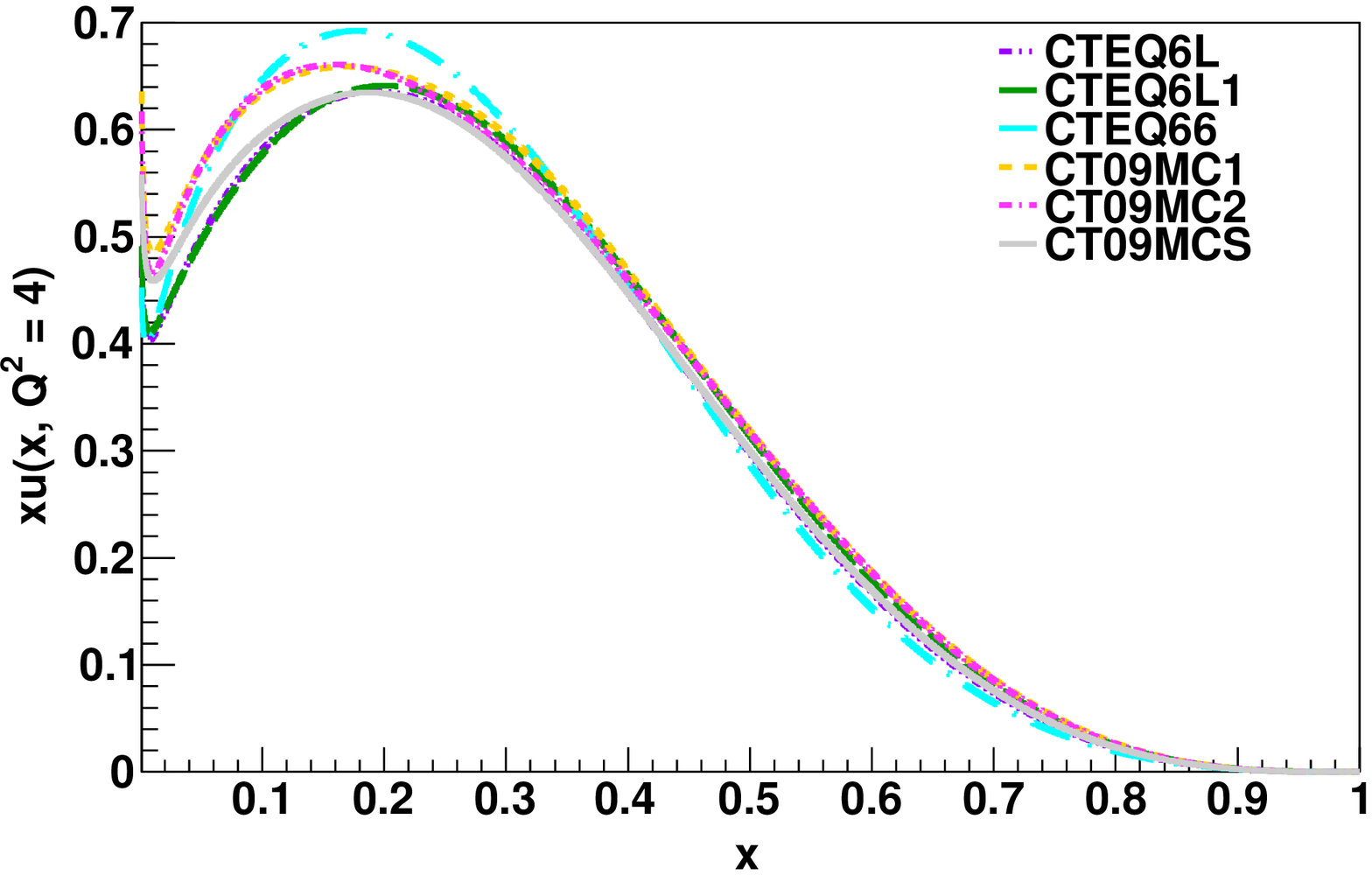}} \\
  \subfloat[]{\label{fig:gLogxQ24-1}\includegraphics[width=0.5\textwidth]
    {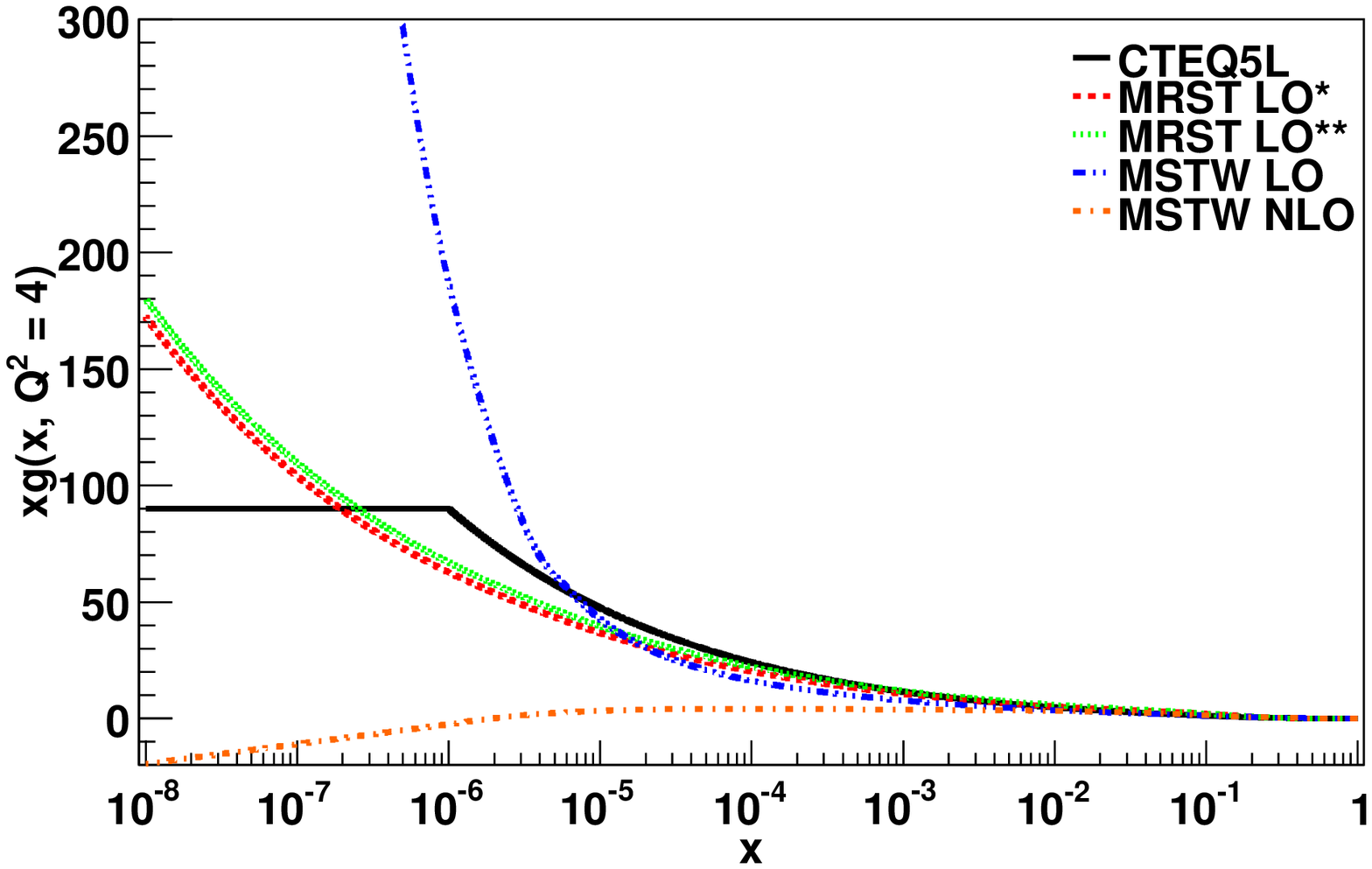}}                
  \subfloat[]{\label{fig:gLogxQ24-2}\includegraphics[width=0.5\textwidth]
    {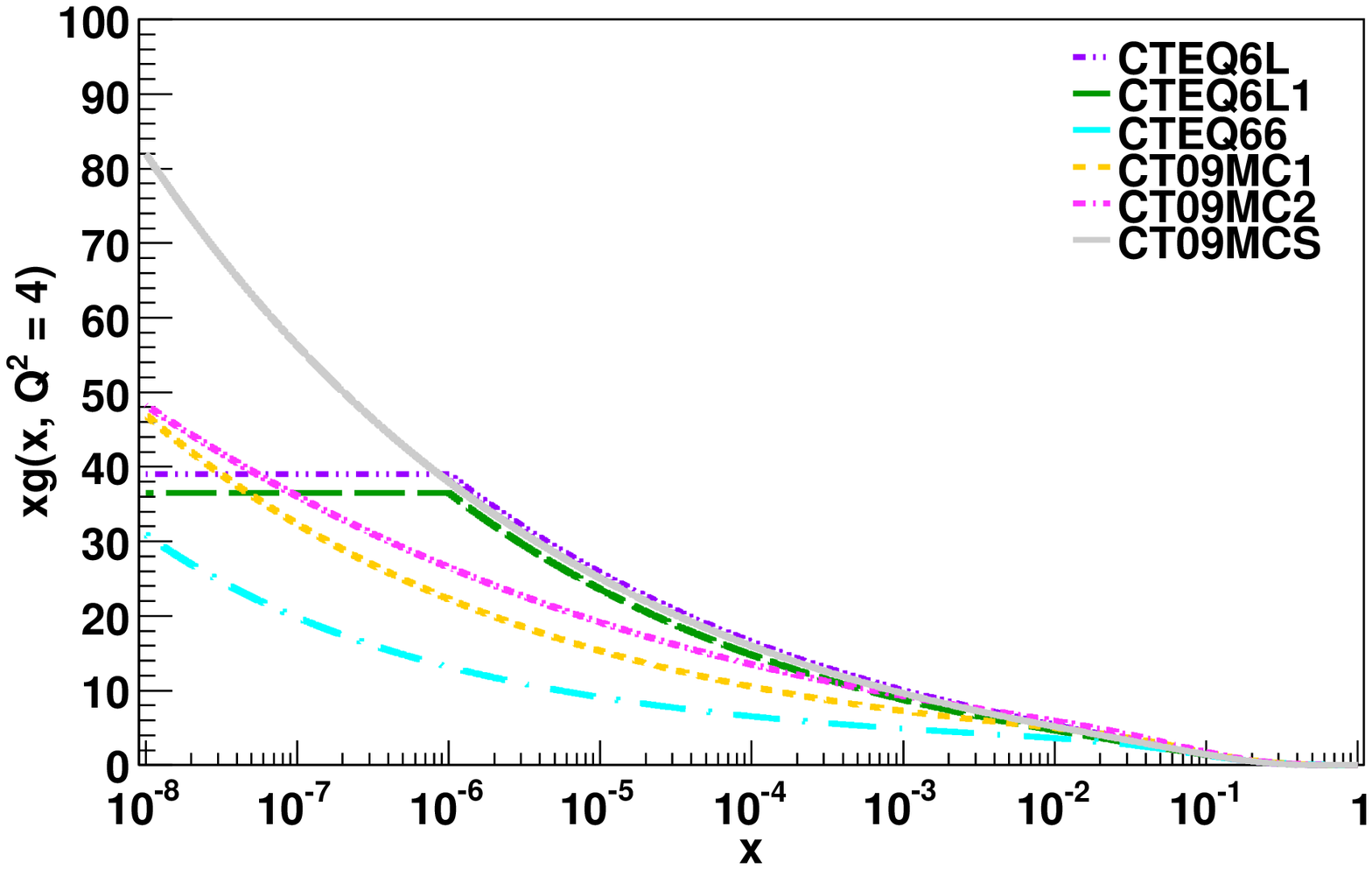}}
  \caption{Up quark (a, b) and gluon (c, d) distributions at
    $Q^2=4$~GeV$^2$. Note the difference in horizontal and vertical scales}
  \label{fig:xfQ24}
\end{figure*}

Minimum-bias events are sensitive to low $Q$; a $Q^2$ around $4$~GeV$^2$ is
a typical scale for such simulations. The up
distributions for large $x$ at $Q^2=4$~GeV$^2$ in Fig.~\ref{fig:xfQ24} are all similar, with slight differences for the two NLO
PDFs and CT09 MC1 and MC2. The gluon distributions, on the other hand, show
large differences in the small-$x$ region: note the difference in horizontal scale. Especially MSTW LO has a much steeper rise and becomes much larger than
the others. All the MRST/MSTW distributions give larger values at small $x$
than the CTEQ ones. The MC-adapted PDFs from the respective collaboration follow each other,
except for MCS which is more similar to CTEQ6L and CTEQ6L1. The two NLO PDFs
stand apart from the rest, and MSTW NLO is negative in a large region. One can also note
that CTEQ5L, CTEQ6L and CTEQ6L1 all freeze at $x=10^{-6}$. 

\begin{figure*}[tp]
  \centering
  \subfloat[]{\label{fig:uxQ21e3-1}\includegraphics[width=0.5\textwidth]
    {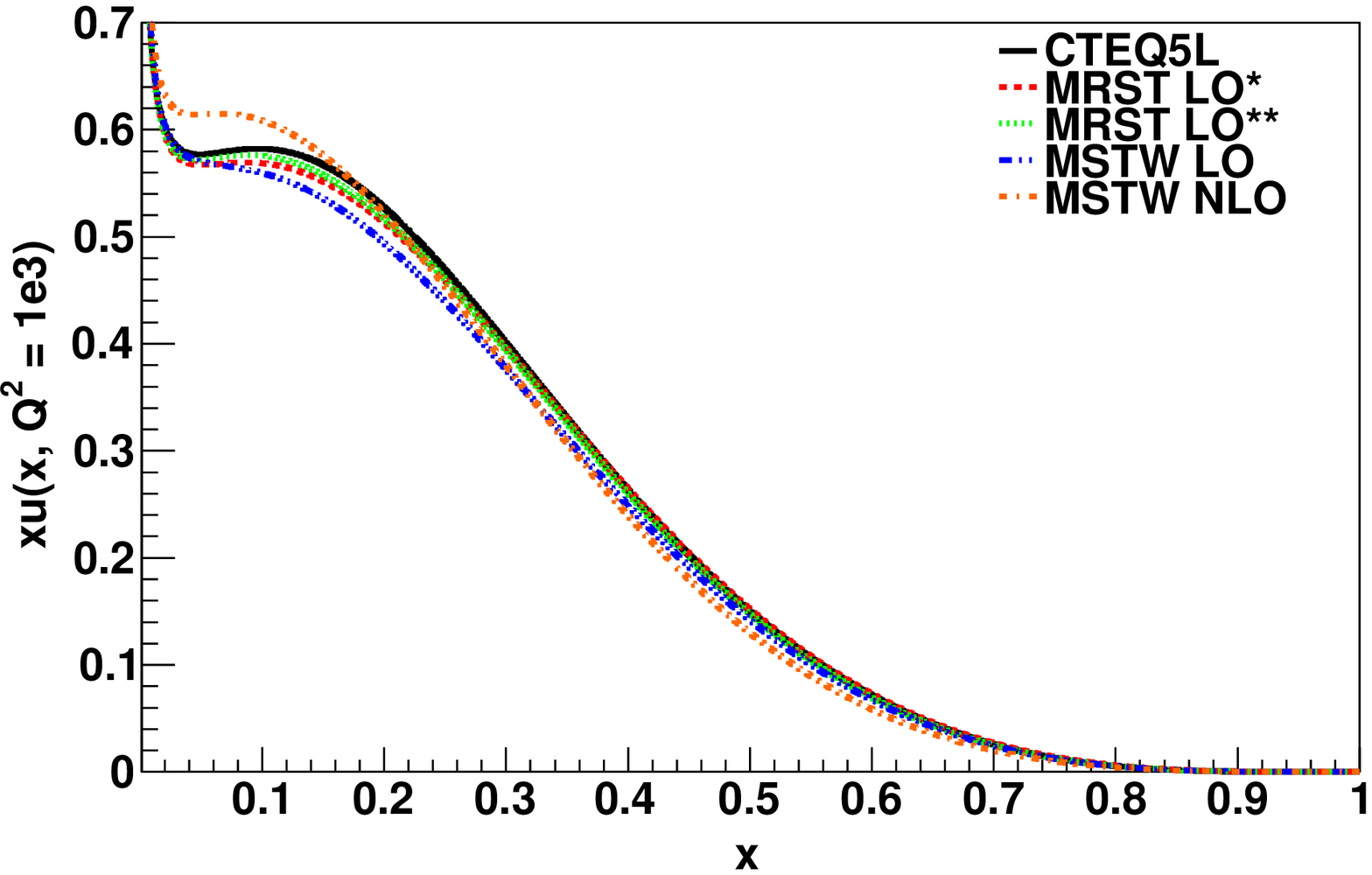}}                
  \subfloat[]{\label{fig:uxQ21e3-2}\includegraphics[width=0.5\textwidth]
    {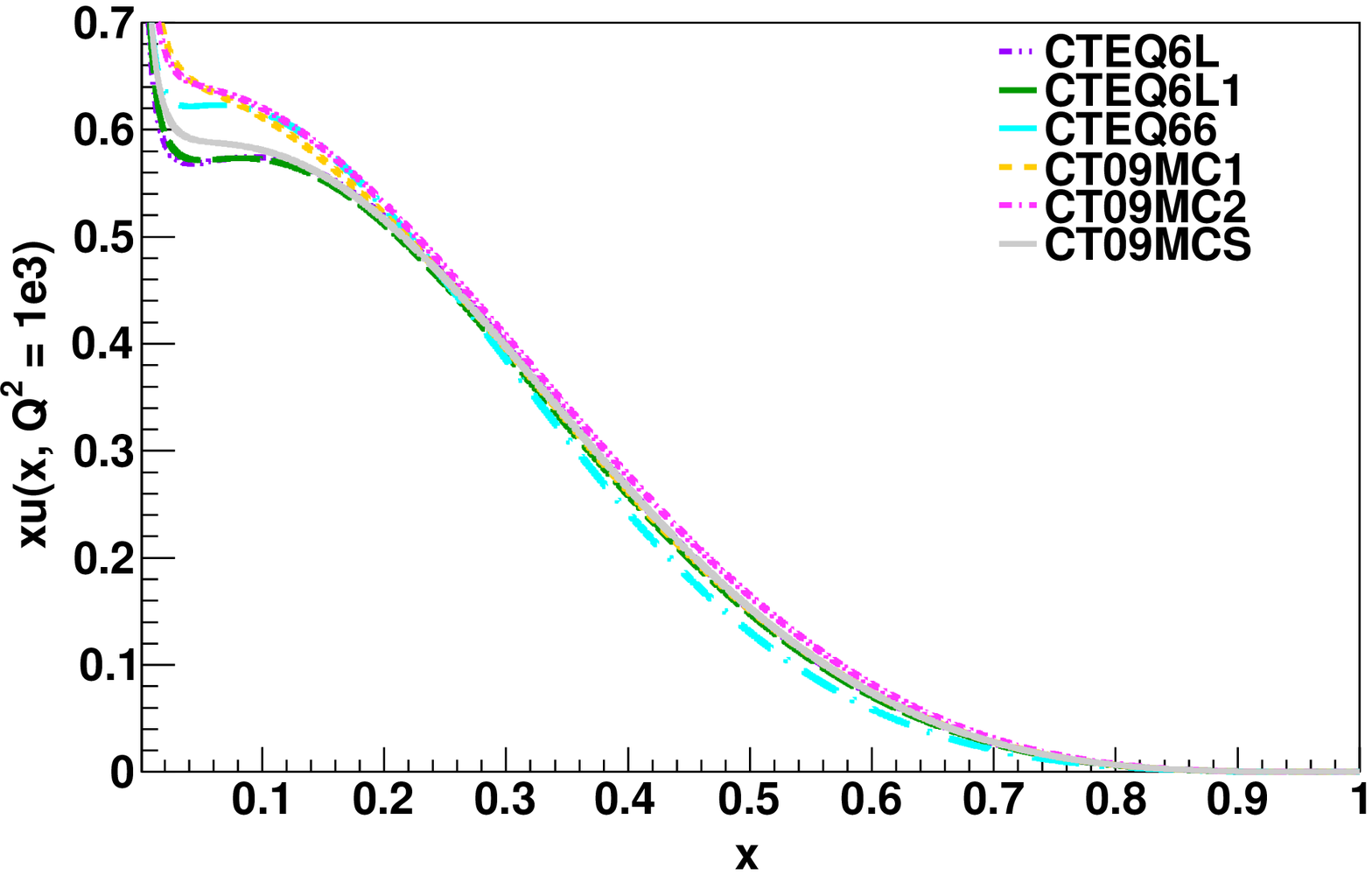}}\\
  \subfloat[]{\label{fig:gLogxQ21e3-1}\includegraphics[width=0.5\textwidth]
    {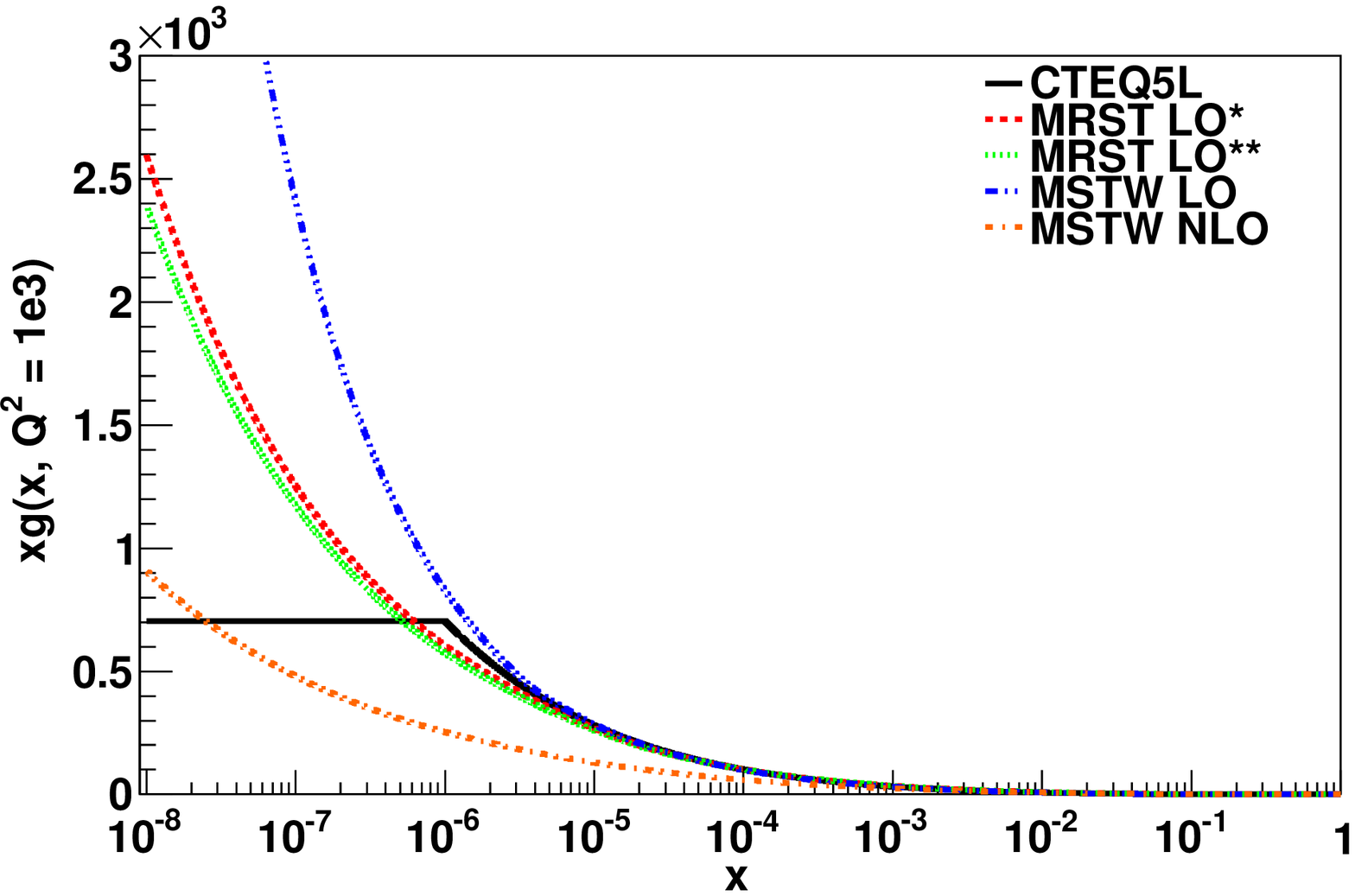}}                
  \subfloat[]{\label{fig:gLogxQ21e3-2}\includegraphics[width=0.5\textwidth]
    {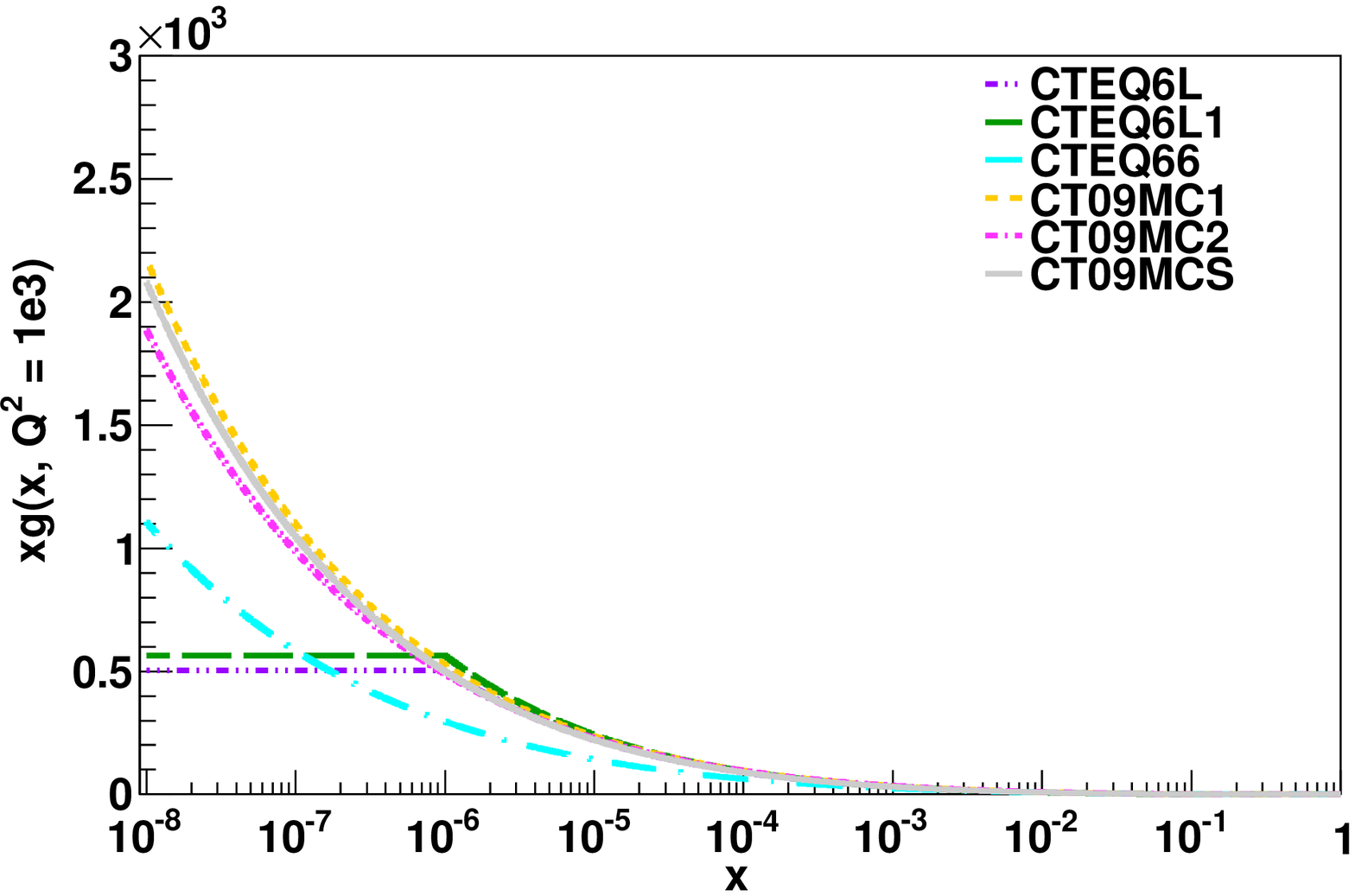}} 
  \caption{Up quark (a, b) and gluon (c, d) distributions at $Q^2=10^{3}$~GeV$^2$. Note the difference in horizontal and vertical scales}
  \label{fig:xfQ21e3}
\end{figure*}

Fig.~\ref{fig:xfQ21e3} shows the distributions at
$Q^2=10^3$. Both the up and the gluon distributions show the
same relative patterns as at $Q^2=4$. However, the differences between the PDFs
are smaller, and especially the difference between the two groups are no
longer as prominent, except where some PDFs freeze. At this $Q^2$ the CTEQ MC-adapted PDFs, in Fig.~\ref{fig:gLogxQ21e3-2}, are all similar.

\section{Minbias Events}
\label{sec:minbias}
 Minbias events are interesting in
their own right, but also because they constitute a background when studying hard
interactions. They tend to
have low
average transverse energy, low particle multiplicity and consist largely of soft
inelastic interactions. Because of the low $p_{\perp}$ the interacting partons only need a
small portion of the momenta of the incoming hadrons, and hence minbias events probe parton distributions in the small-$x$ region dominated by the gluon distribution.

We examine the rapidity and multiplicity distributions from
simulations with different PDFs, both at Tevatron and LHC energies. With the aid of Rivet \cite{Rivet} we also
compare $p_{\perp}$ and $\sum E_{\perp}$ particle spectra, as well as average
$p_{\perp}$ evolution with
multiplicity, with data collected by the CDF experiment at Tevatron Run 2 \cite{Aaltonen2}.

\subsection{Multiplicity and Tuning}
A generator using the different PDFs needs to be retuned separately for each of them before we can make any reasonable comparisons. Specifically, the larger momentum carried by the partons in LO*, LO**, MC1 and MC2 allows a larger activity and hence a larger multiplicity than with the
ordinary leading-order PDFs. Furthermore the NLO PDFs give less activity,
owing to the small gluon at low $x$ and $Q^2$. We have chosen to tune \textsc{Pythia8} so that all PDFs 
have the same average charged-particle multiplicity as CTEQ5L. This PDF
is taken as reference because it is the default PDF in \textsc{Pythia8} and is
most commonly used in \textsc{Pythia8} simulations. We are not making a complete tune
and only intend to get a first impression of relative differences, under
comparable conditions. The tuning is accomplished by tweaking the
$p_{\perp 0}^{\stext{Ref}}$ parameter in \textsc{Pythia8}
\begin{equation}
  p_{\perp 0}=p_{\perp 0}^{\stext{Ref}}\left( \frac{E_{\stext{CM}}}{E_{\stext{CM}}^{\stext{Ref}}}\right)^p
\end{equation}
where $E_{\stext{CM}}^{\stext{Ref}} = 1800$~GeV and $p = 0.24$. $p_{\perp 0}$ is used for the regularization of the divergence of the QCD
cross section as $p_{\perp}\rightarrow 0$. A smaller $p_{\perp0}$ cause the
regularization to take effect at a lower $p_{\perp}$, increasing the charged
particle multiplicity, $n_{ch}$. The tuning
was done for the $\alpha_S$ value and leading-order running which is default
in \textsc{Pythia8}. Results are shown in Tab.~\ref{charmult1}. The largest multiplicity before retuning is
obtained with LO**, followed by MC2 and LO* which are also the PDFs with
most momentum. 

Further tuning could well bring some of the results closer together, but 
it is well known that $p_{\perp 0}^{\stext{Ref}}$ is the single most crucial
parameter for obtaining overall agreement with minimum-bias data at a 
given energy. It is also a parameter without any constraints from theory.
Other key min-bias parameters, related to the impact-parameter picture, the 
colour-reconnection mechanism or the beam-remnant handling, do not have 
as direct a coupling to the choice of PDF.  
The parton showers, by contrast, are better controlled by basic principles. 
Even if there are different shower schemes being proposed and used, within 
a given scheme there is no degree of freedom intended to counteract the 
PDF (and associated $\alpha_S$) choice, in the way that 
$p_{\perp 0}^{\stext{Ref}}$ is. The exception would be $K$ factors for 
cross sections, which we will comment on later. Hadronization, finally, 
is fixed by the LEP data, and should not be touched.

\begin{table}
		\center
    \begin{tabular}{lcc}
    \toprule
    PDF & $\langle n_{ch} \rangle$ & $p_{\perp 0}^{\stext{Ref}}$ \\
    \midrule
    CTEQ5L    & 54.48 & 2.25 \\
    MRST LO*  & 59.74 & 2.50 \\
    MRST LO** & 63.52 & 2.63 \\
    MSTW LO   & 49.10 & 2.06 \\
    MSTW NLO  & 48.02 & 1.56 \\
    CTEQ6L    & 54.92 & 2.25 \\
    CTEQ6L1   & 51.71 & 2.13 \\
    CTEQ66    & 42.85 & 1.75 \\
    CT09MC1   & 53.92 & 2.25 \\
    CT09MC2   & 60.37 & 2.50 \\
    CT09MCS   & 54.87 & 2.25 \\
    \bottomrule
   \end{tabular}
   \caption{Average charged particle multiplicity for the different PDFs with the default value of
    $p_{\perp 0}^{\stext{Ref}} = 2.25$ and also the $p_{\perp 0}^{\stext{Ref}}$ required to tune the charge
    multiplicity to be equal to the value for CTEQ5L}
   \label{charmult1}
\end{table}

\subsection{Generator-level}
\begin{figure*}[tp]
  \subfloat[]{\label{fig:CRapid-allstrip1}\includegraphics[width=0.5\textwidth]
    {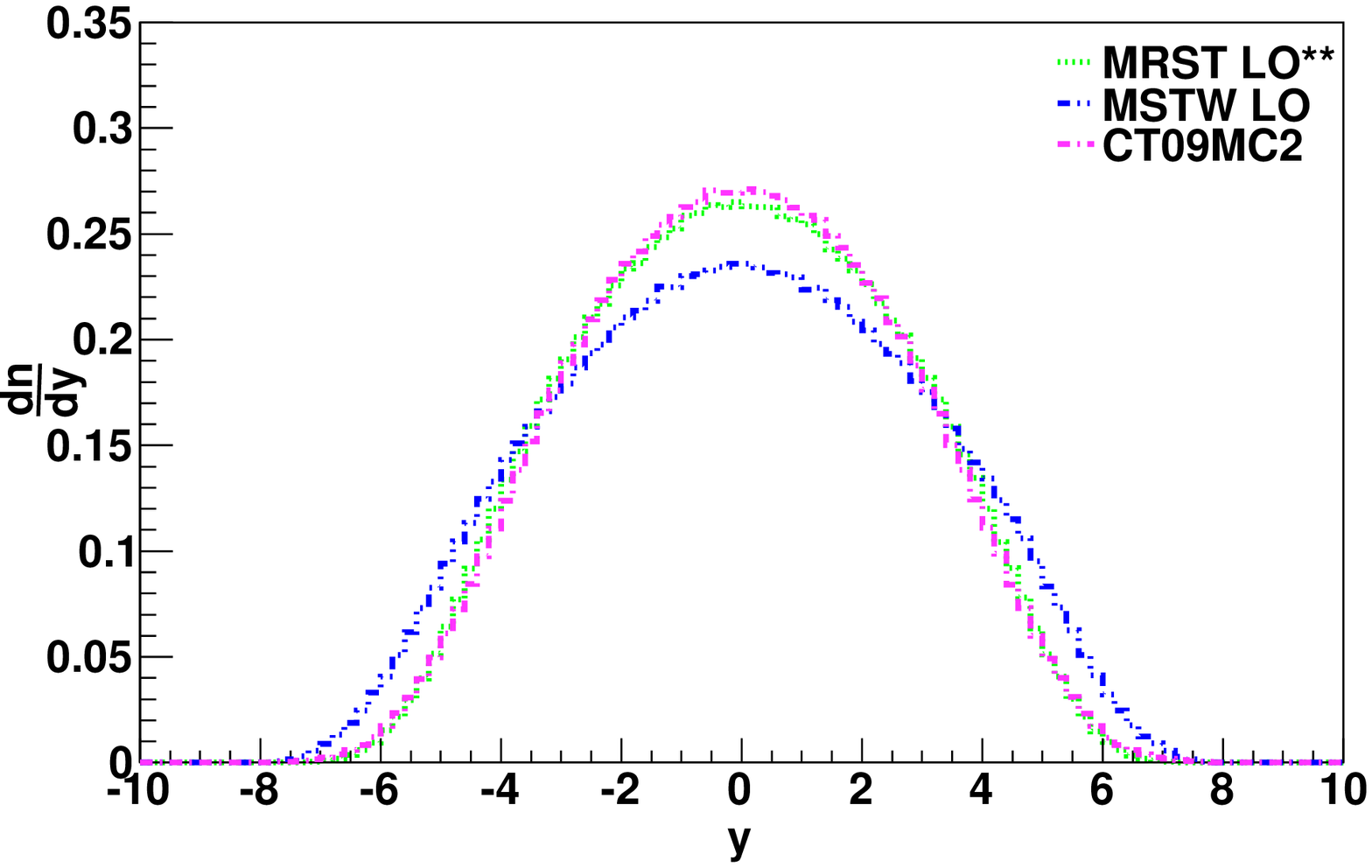}}
   \subfloat[]{\label{fig:CRapid-allstrip2}\includegraphics[width=0.5\textwidth]
    {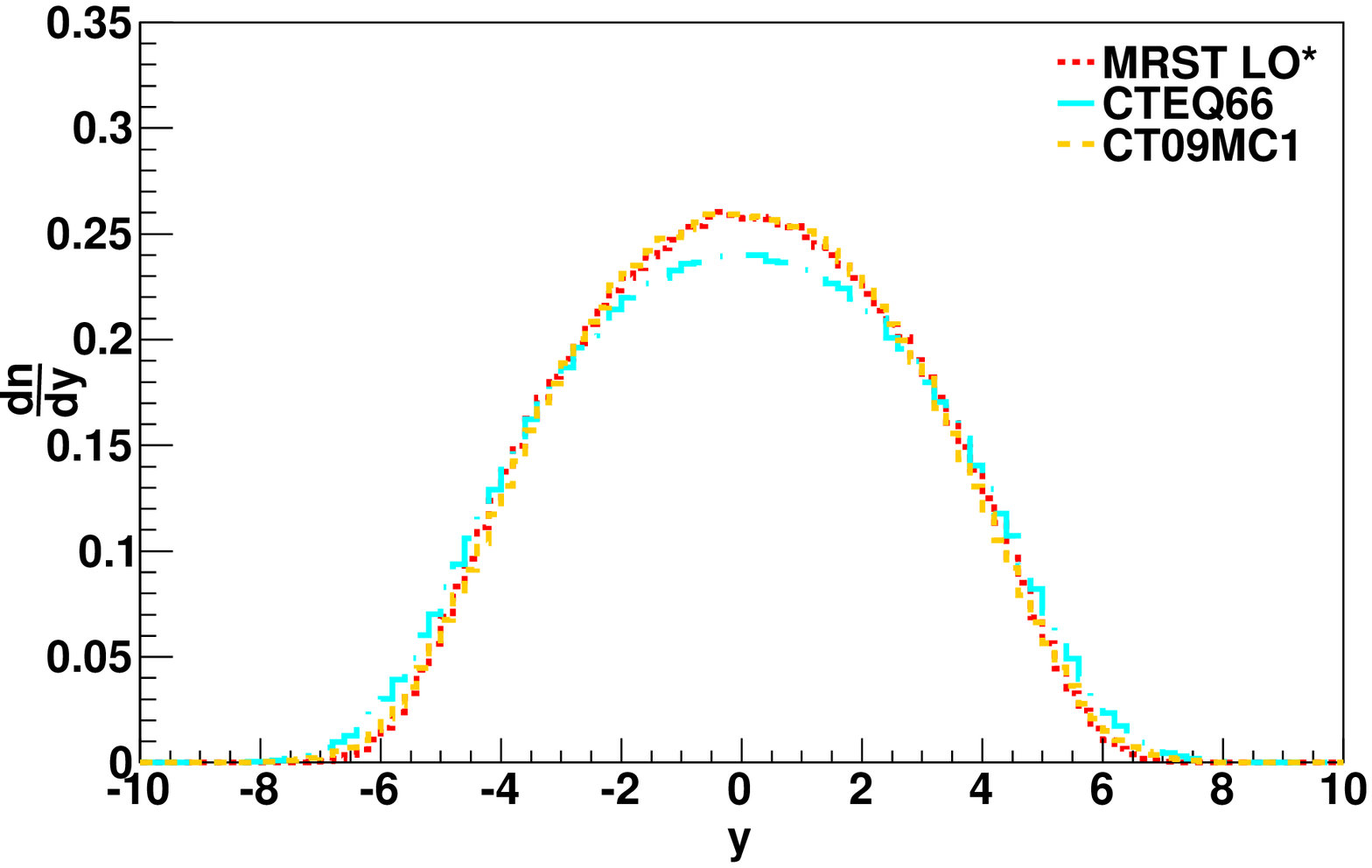}}
  \caption{Rapidity distributions of the partons created in the $2 \rightarrow 2$ sub-process}
  \label{fig:Rapid-strip}
\end{figure*}

In this section all simulations are for $p\bar{p}$ collisions at 1960~GeV, if not
explicitly stated otherwise. So as not to crowd the plots, not all sets are
shown all the time. The selection of PDFs that are shown each figure represents both
the extremes and the middle way. The
rapidity distributions of the outgoing particles at the parton level, when only
the $2 \rightarrow 2$ sub-process is considered, are presented in
Fig.~\ref{fig:Rapid-strip}. MSTW LO gives a broader distribution than the rest of the PDFs as an effect of the large gluon
distribution at small $x$. The distributions with LO** and MC2 closely resemble
each other, as do the distributions with LO* and MC1. This may be explained by the similarities in the violation of the momentum sum rule. The NLO PDFs give lower
values in the central
rapidity region. The remaining leading-order PDFs give
results which are all similar to the MC-adapted ones.

\begin{figure*}[tp]
  \subfloat[]{\label{fig:CRapid1}\includegraphics[width=0.5\textwidth]
    {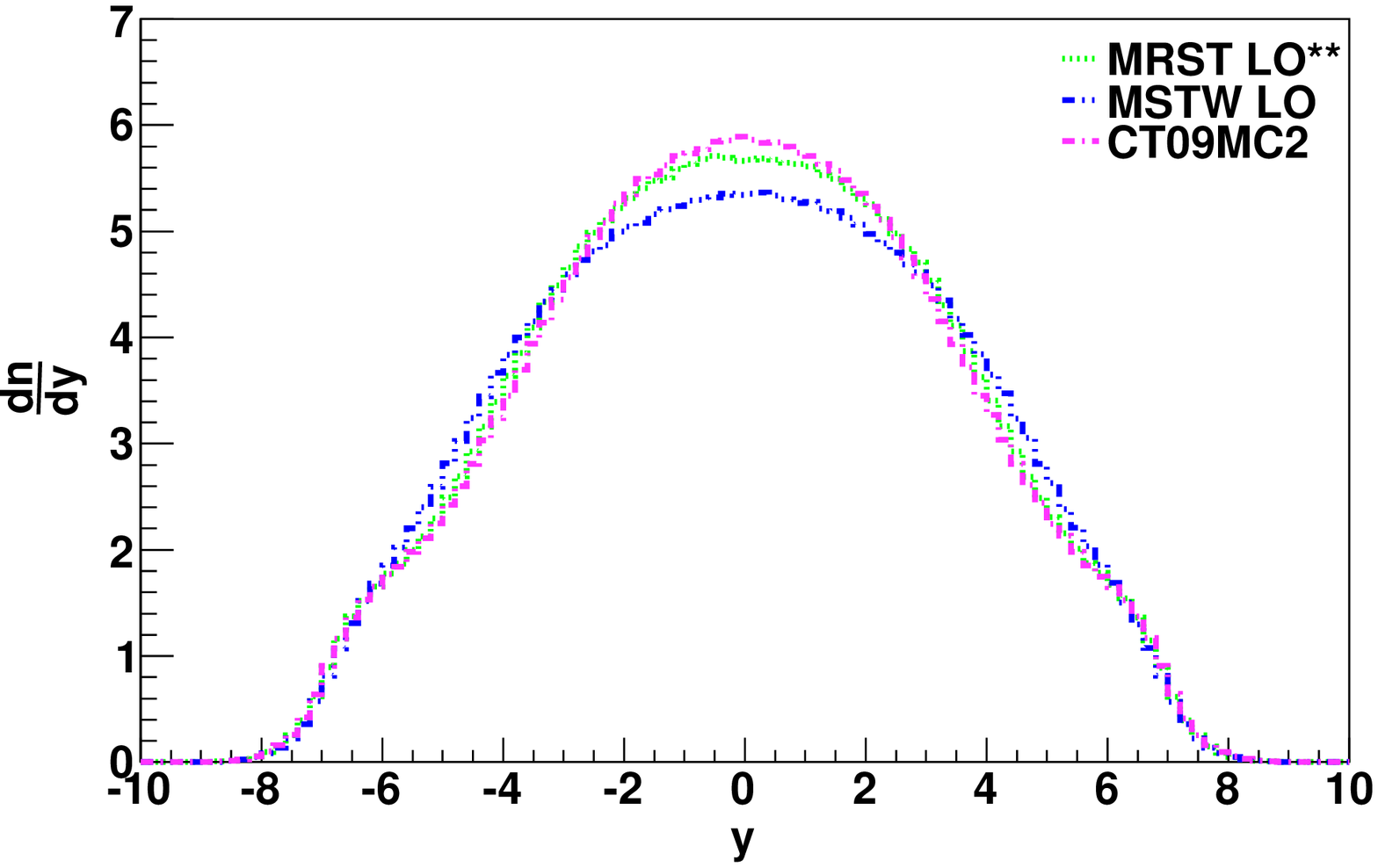}}
   \subfloat[]{\label{fig:CRapid2}\includegraphics[width=0.5\textwidth]
    {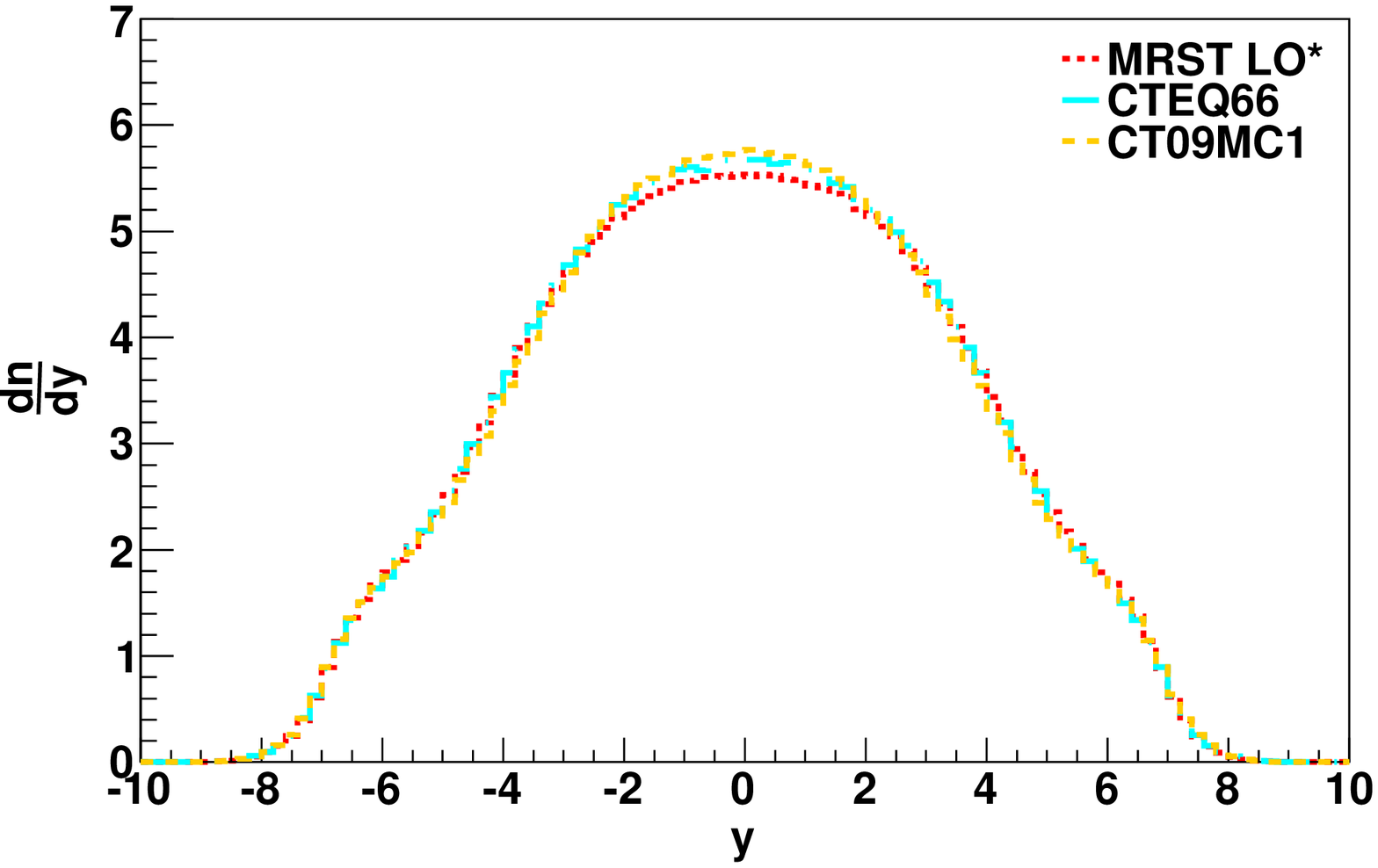}}
  \caption{Rapidity distributions of charged particles after hadronization at
    minbias simulations with $E_{CM}=1960$~GeV}
  \label{fig:Rapid}
\end{figure*}

Turning on the
rest of the \textsc{Pythia8} machinery, the distribution of charged
particles after hadronization, shown in Fig.~\ref{fig:Rapid}, 
changes in shape. There are now more particles at larger rapidities as a
result of the fragmenting color field strings stretched out to the beam remnants, and most of the differences
between the PDFs get blurred. Some differences still remain. MSTW LO still gives smaller multiplicity at central rapidities, and as a remnant of the wider
distribution lack the inward dents that all other PDFs give at rapidities
around $\pm5$. The peaks for LO** and MC2 are somewhat sharper than for LO*
and MC1, but the trace of the lower value with NLO PDFs at central rapidity has vanished.

\begin{figure*}[tp]
  \subfloat[]{\includegraphics[width=0.5\textwidth]
    {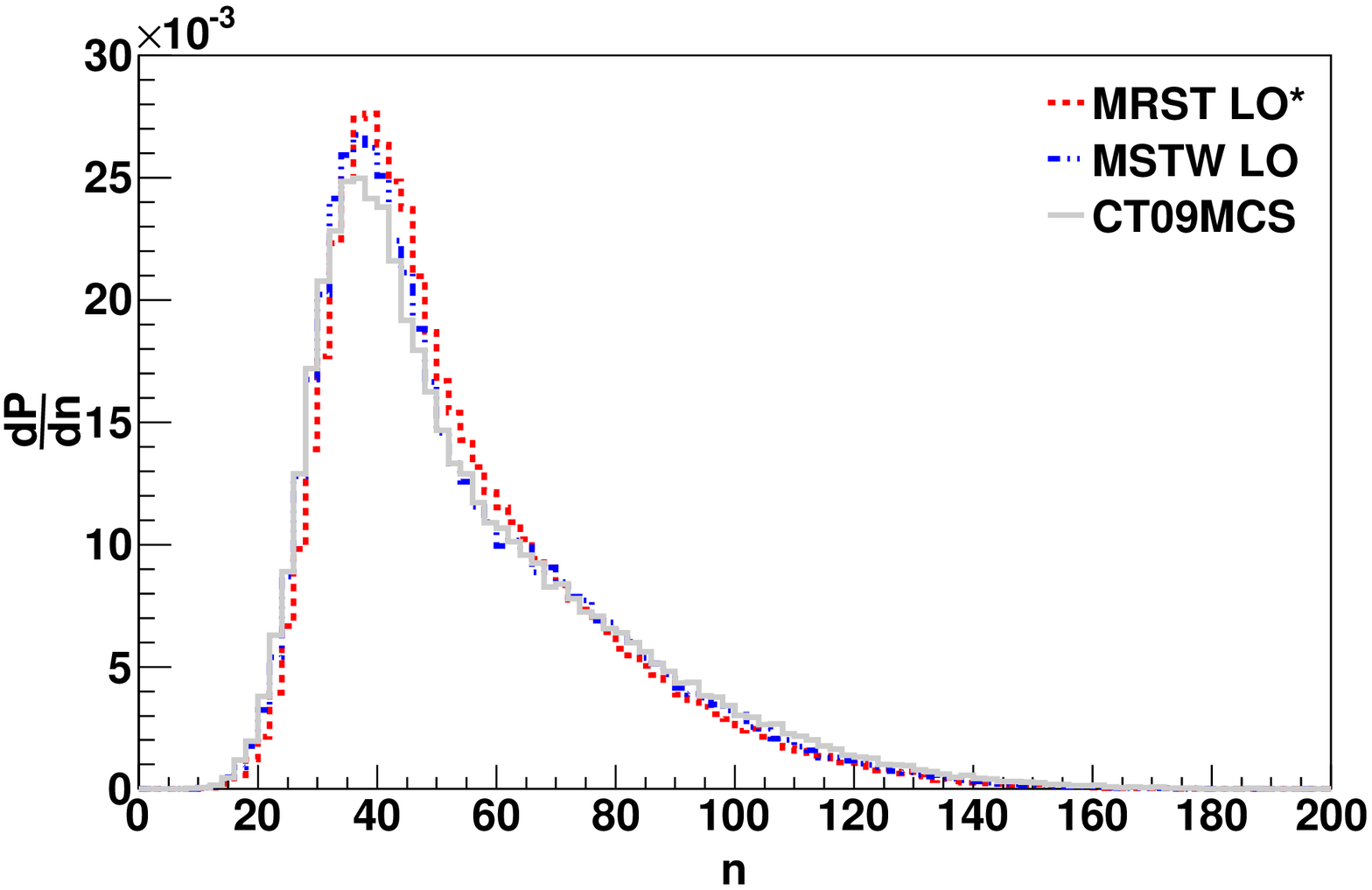}}
   \subfloat[]{\includegraphics[width=0.5\textwidth]
    {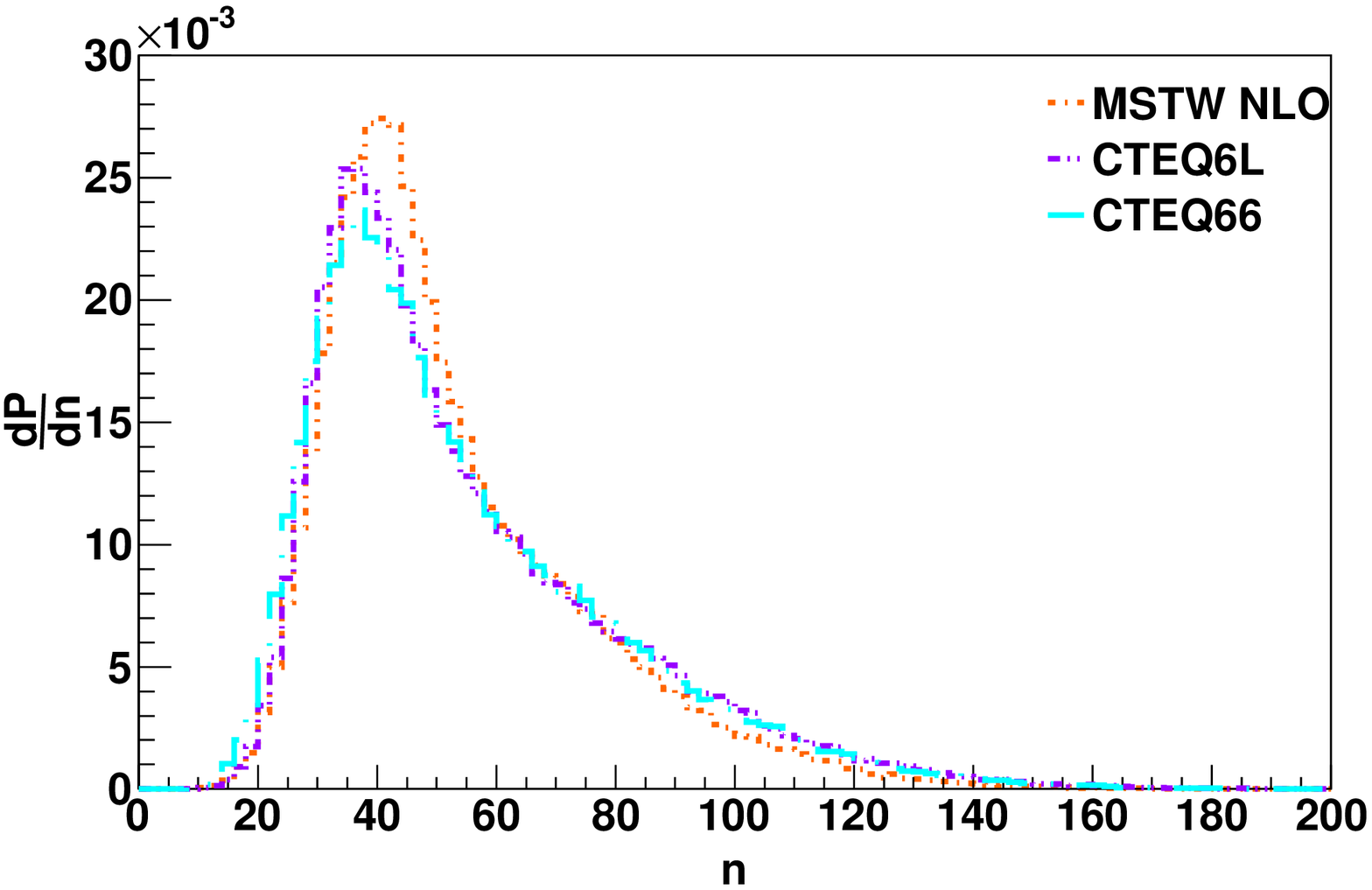}}
  \caption{Charged particle multiplicity distributions}
  \label{fig:Multdist}
\end{figure*}

The multiplicity distributions are similar for most
PDFs. The two NLO distributions stand out as two extremes in different
directions, MSTW NLO here gives the highest peak and the shortest tail as shown
in Fig.~\ref{fig:Multdist}. All MC-adapted PDFs, except MCS,
give a peak slightly shifted to larger multiplicities but are different in
height, where the two from MRST give a larger peak value. The three normal
leading-order distributions are all similar and we only show the CTEQ6L.

\begin{figure*}[tp]
  \subfloat[]{\includegraphics[width=0.5\textwidth]
    {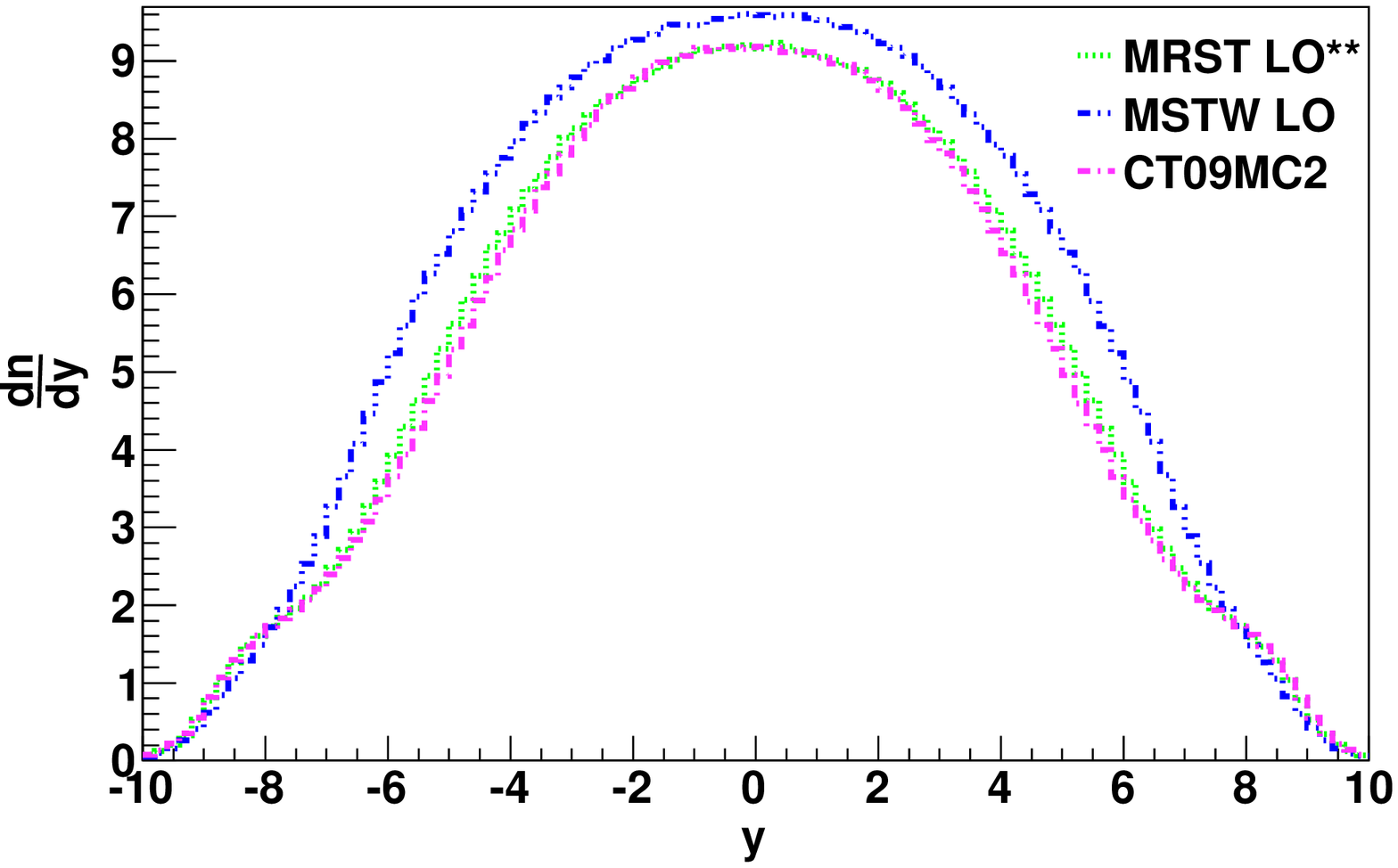}}
   \subfloat[]{\includegraphics[width=0.5\textwidth]
    {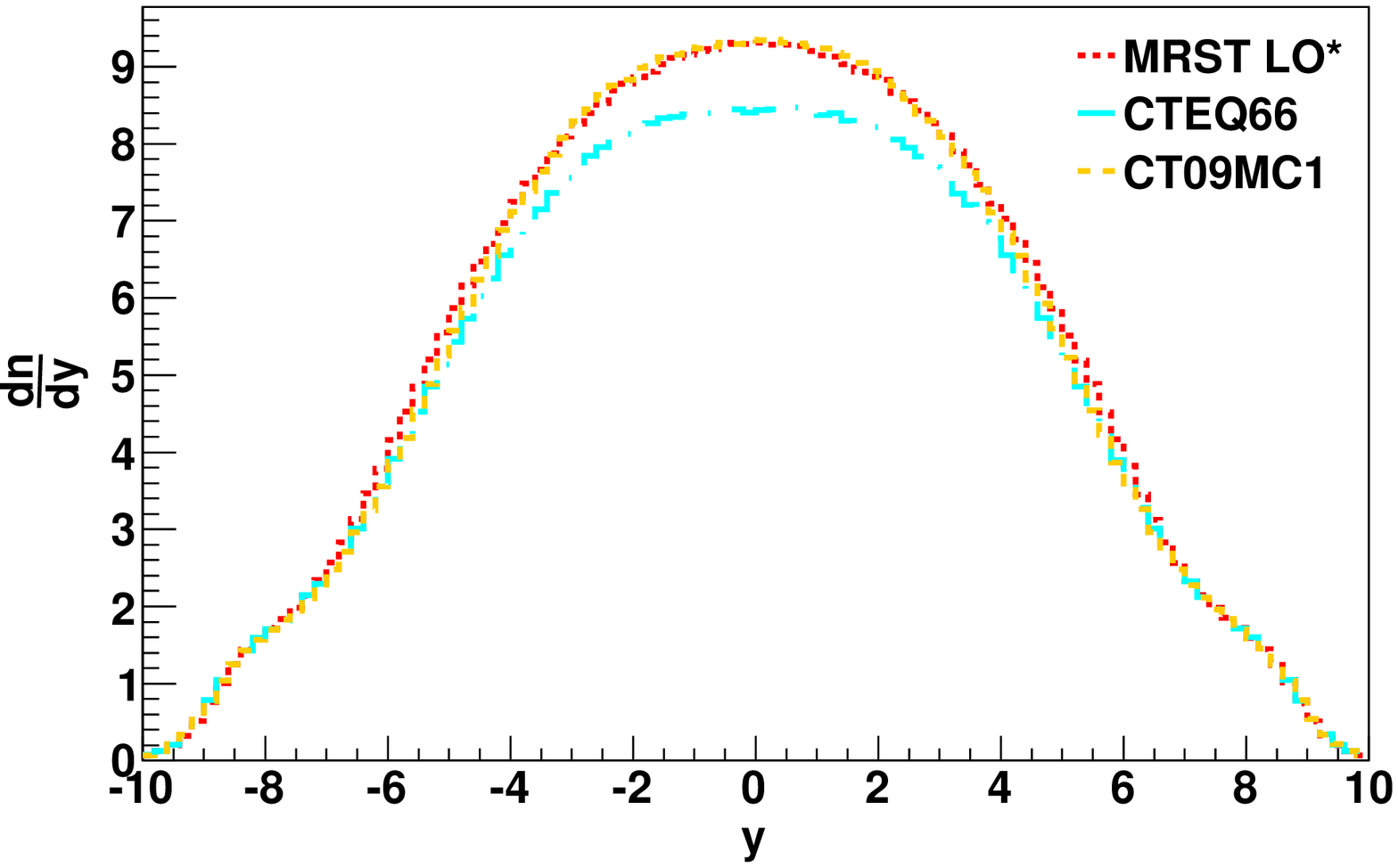}}
  \caption{Rapidity distributions at LHC ($pp$ collisions at $E_{CM}=14$~TeV)}
  \label{fig:Rapid-LHC}
\end{figure*}

Increasing the energy
to the level of a fully operational LHC enhances the differences
seen at Tevatron energy, especially for MSTW LO and the two NLO PDFs. The
multiplicity of these three evolve with energy in a different way than for the
other PDFs. The rapidity distribution, shown in Fig.~\ref{fig:Rapid-LHC},
naturally extends to larger rapidities and shows a higher central activity. MSTW LO here gives
a much broader distribution. This is because as the energy increases even lower values of $x$
come into play, so that the effect of the gluon distribution in this region
has larger impact on the results. The two NLO PDFs result in flatter peaks
than the MC-adapted PDFs and are similar in shape to MSTW LO. This can be explained by the smaller gluon distribution at small $x$. The rest of the PDFs evolve in a fashion similar to the
MC-adapted PDFs shown in the figure, but with some more variation. MC1 and LO* results in a little bit larger distributions at central rapidities than MC2 and LO**.

\begin{figure*}[tp]
  \subfloat[]{\includegraphics[width=0.5\textwidth]
    {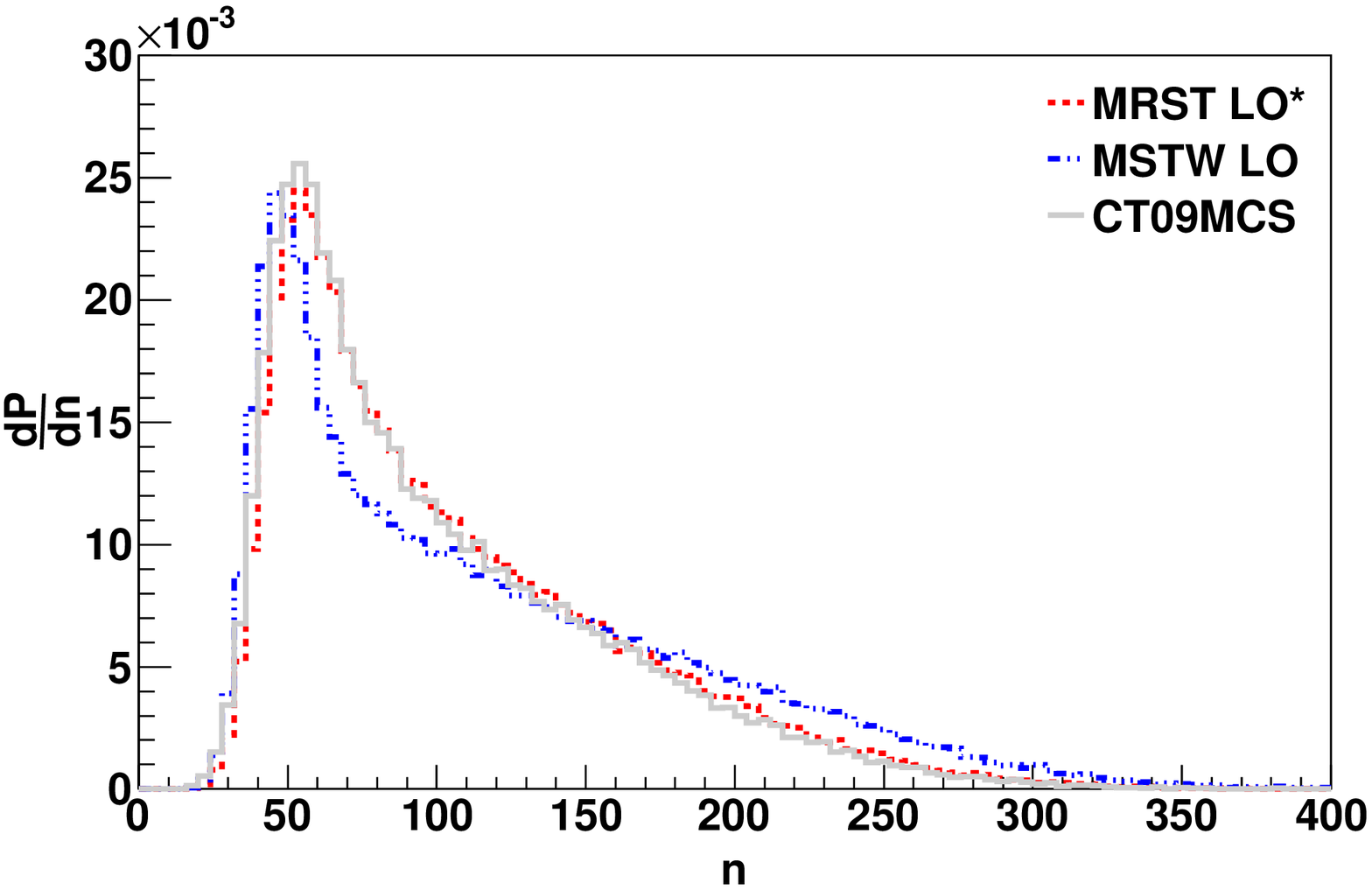}}
   \subfloat[]{\includegraphics[width=0.5\textwidth]
    {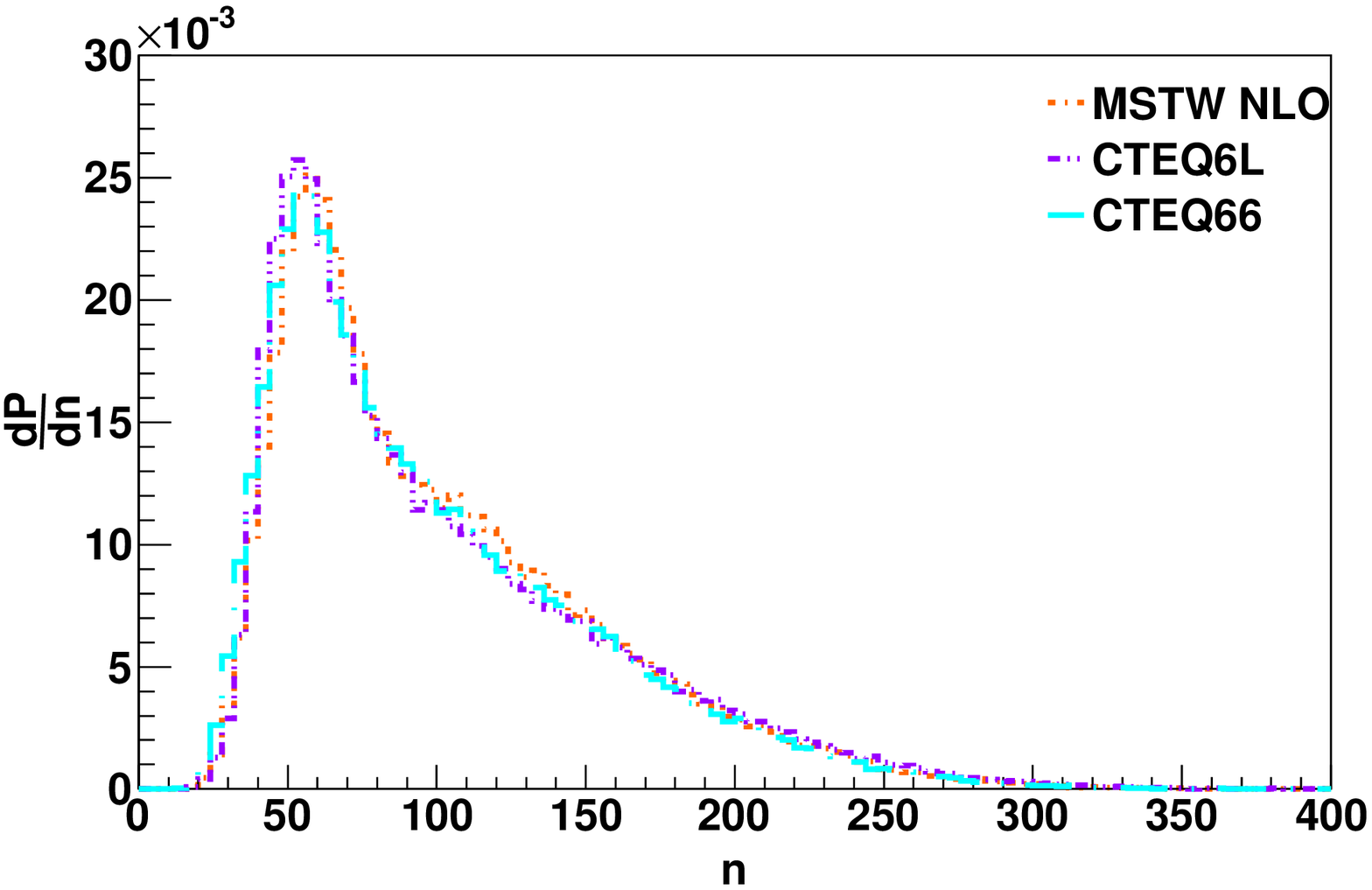}}
  \caption{Charged particle multiplicity distributions at LHC}
  \label{fig:Mult-LHC}
\end{figure*}

The multiplicity distributions in Fig.~\ref{fig:Mult-LHC} also show the effect of the low-$x$ gluon distribution. MSTW LO has a much higher total multiplicity. The two NLO PDFs converge at this energy but they have smaller multiplicity than MSTW LO because of their small gluon distribution at small $x$. In general the same distributions that stand out with their rapidities also do so with their charge particle multiplicity
distributions.

\subsection{Comparison with CDF Run 2 data}
\begin{figure*}[tp]
  \centering
  \subfloat[]{\includegraphics[width=0.5\textwidth]
    {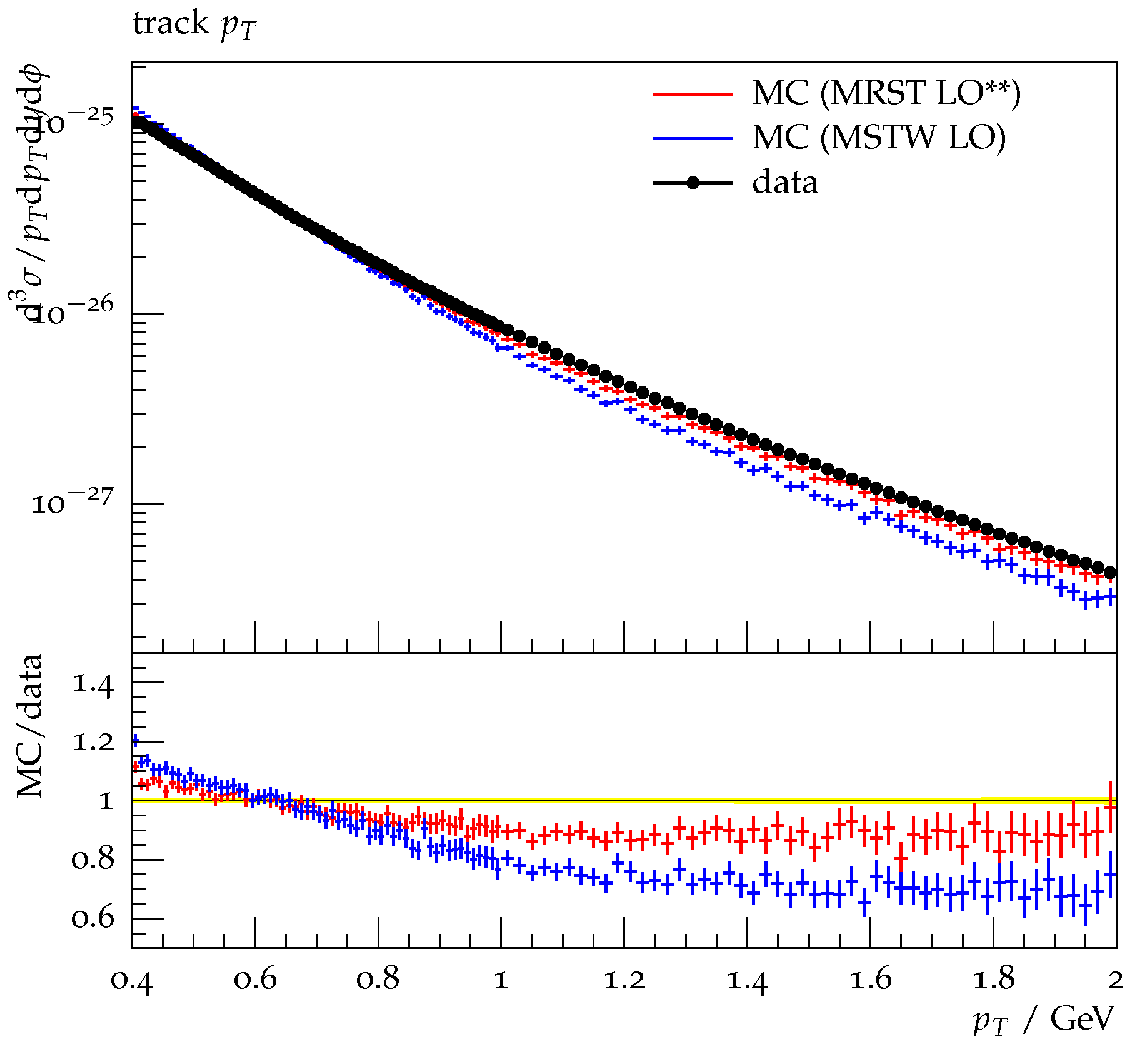}} 
  \subfloat[]{\includegraphics[width=0.5\textwidth]
    {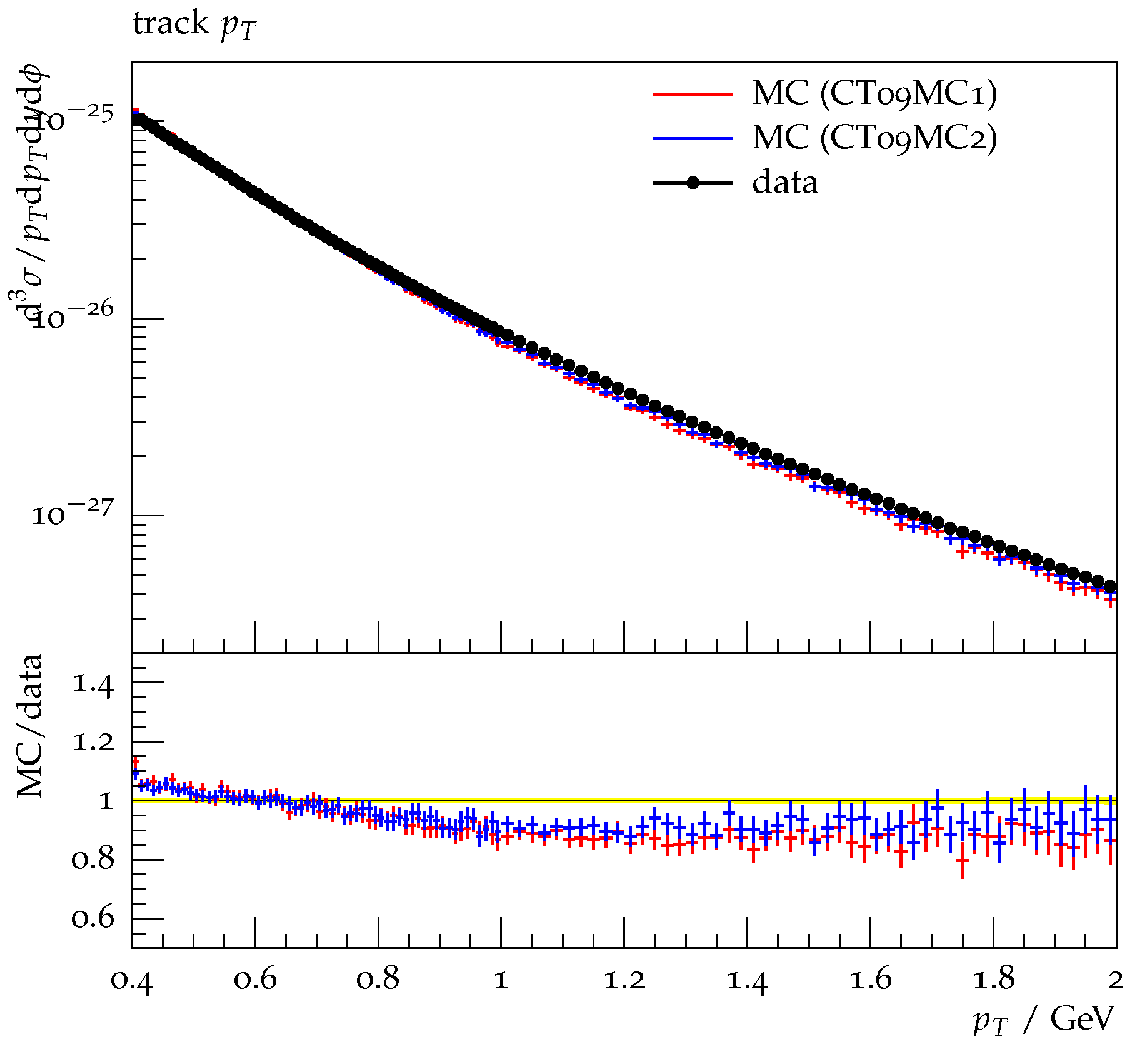}}
  \caption{$p_{\perp}$ spectra of charged particles from the CDF Run 2 experiment compared with
    simulations with different PDFs}
  \label{fig:SigmapT-Riv}
\end{figure*}

The analyses in Rivet ensure that the comparisons to data have the same cuts
and corrections as the original experiment. Therefore only the central
pseudorapidity region is used and cuts in transverse
momentum are implemented \cite{Aaltonen2}. 
$p_{\perp}$ spectra of charged particles in Fig.~\ref{fig:SigmapT-Riv} show the same
relative shape for all PDFs, which gives too large values at the 
low-$p_{\perp}$ end,
then decreases compared to data and gives too small differential cross sections
at the high end. The slope shows some differences depending on
the choice of PDF. MC-adapted PDFs and the CTEQ6L give results that are the
closest to data, while MSTW LO and NLO are further away than the rest.

The $\sum E_{\perp}$ spectrum of particles, neutral particles included, shows
larger dependence on the PDFs, but the trend of disagreement of the PDFs with the data is the same as that noted for $p_{\perp}$. Since we have not done a complete tune,
differences 
do indicate the importance of the PDFs. Further tuning could well bring curves
closer to each other, but the $\sum E_{\perp}$ distribution is less dependent on
details of the MC and therefore easier for PDF developers to consider in tunes. 
The MC-adapted PDFs reproduce data well while the LO and NLO PDFs give results which are further away. The exception is CTEQ6L which also gives results close to data, while MSTW LO goes down to less than half the cross
section of data at the larger energy end. MSTW NLO results peak at higher energies
than the rest. All PDFs give a too large value at the peak, but then decrease
too fast and differ the most from data at the high energy end.

\begin{figure*}[tp]
  \centering
  \subfloat[]{\includegraphics[width=0.5\textwidth]
    {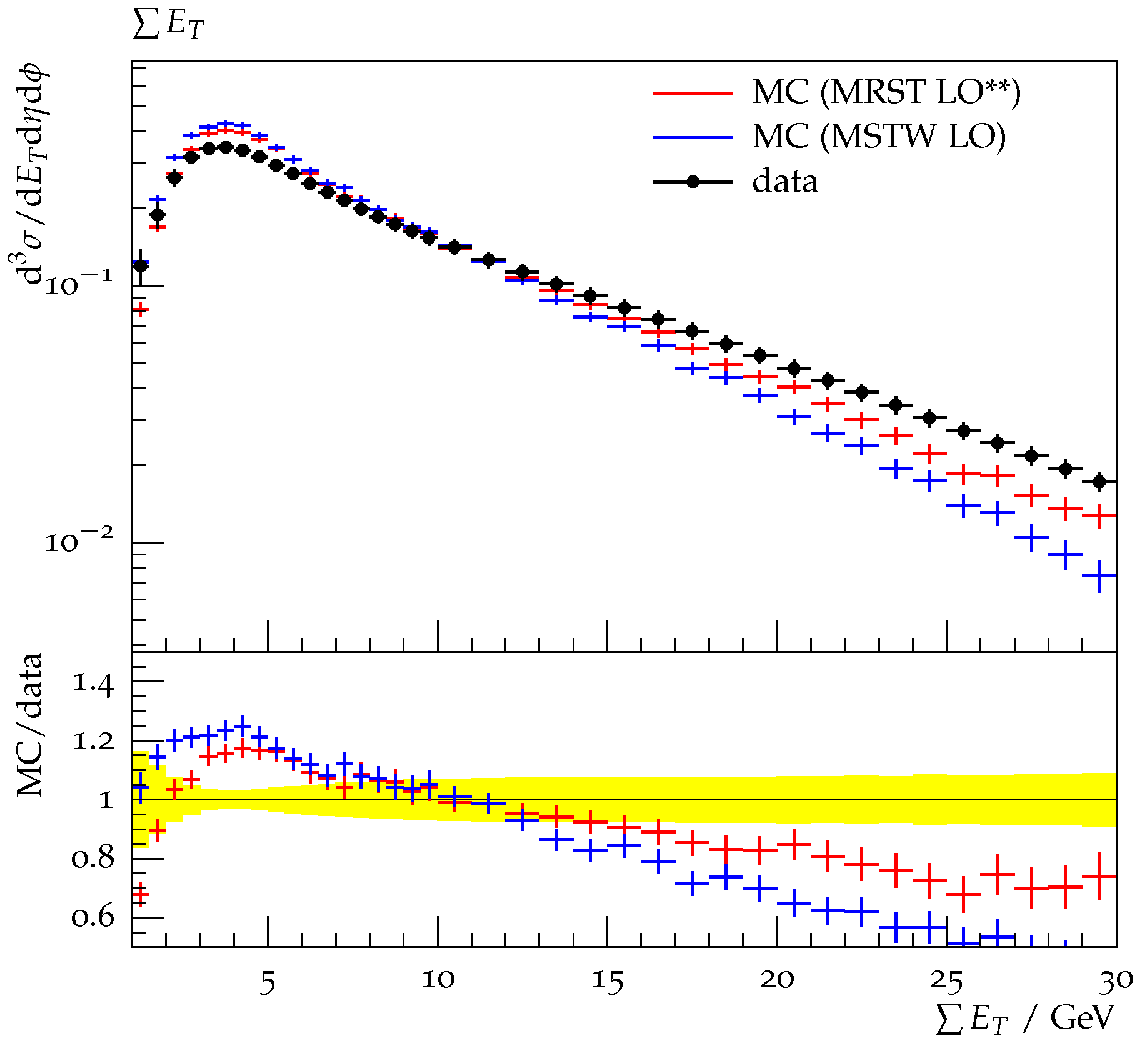}} 
  \subfloat[]{\includegraphics[width=0.5\textwidth]
    {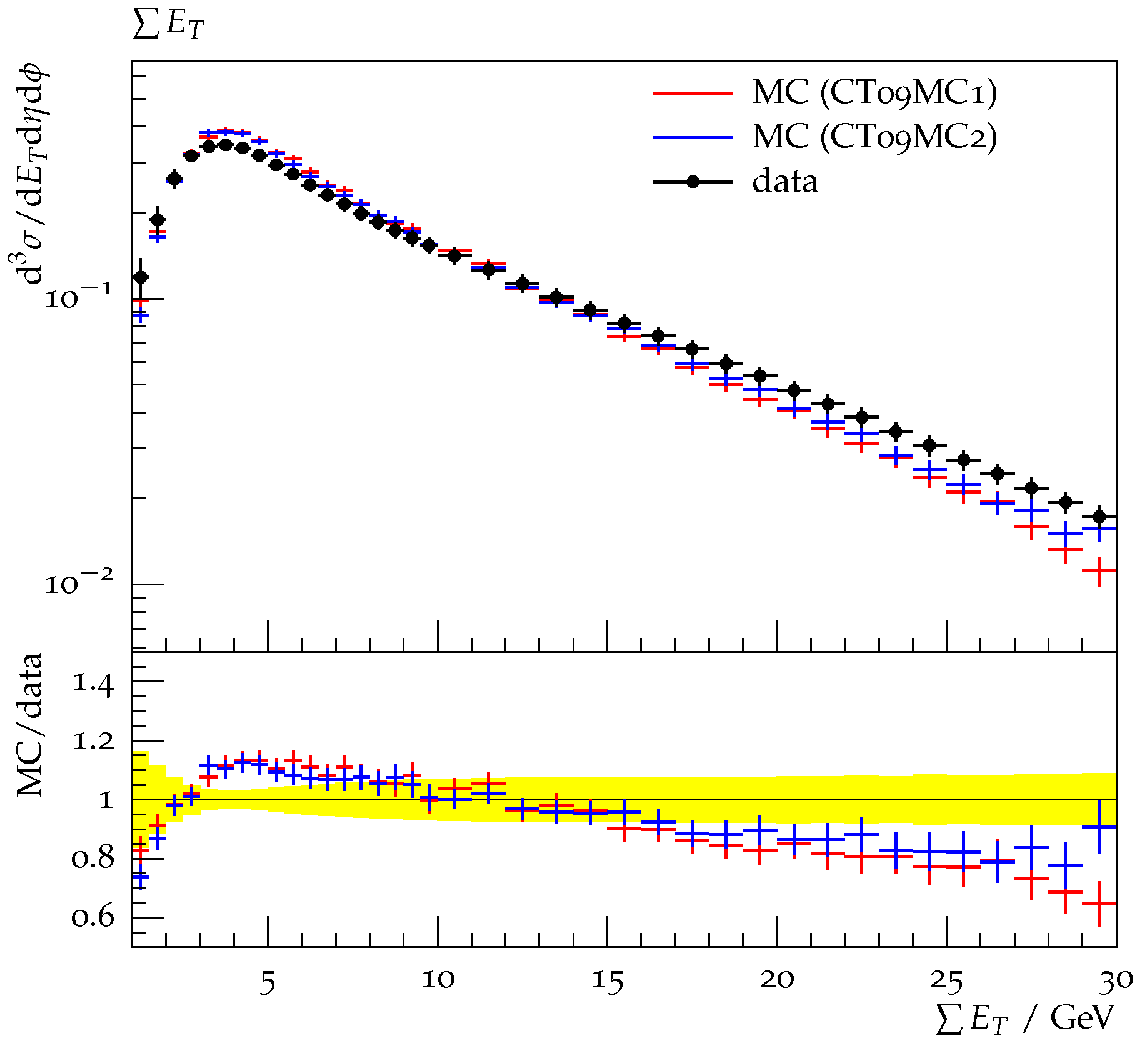}}
\vspace{-0.2cm}
  \caption{$\sum E_{\perp}$ spectra from the CDF Run 2 experiment compared with
    simulations with different PDFs}
  \label{fig:SigmaET-Riv}
\vspace{-0.4cm}
\end{figure*}

\begin{figure*}[tp]
  \centering
  \subfloat[]{\label{fig:pTCMult-Riv1}\includegraphics[width=0.5\textwidth]
    {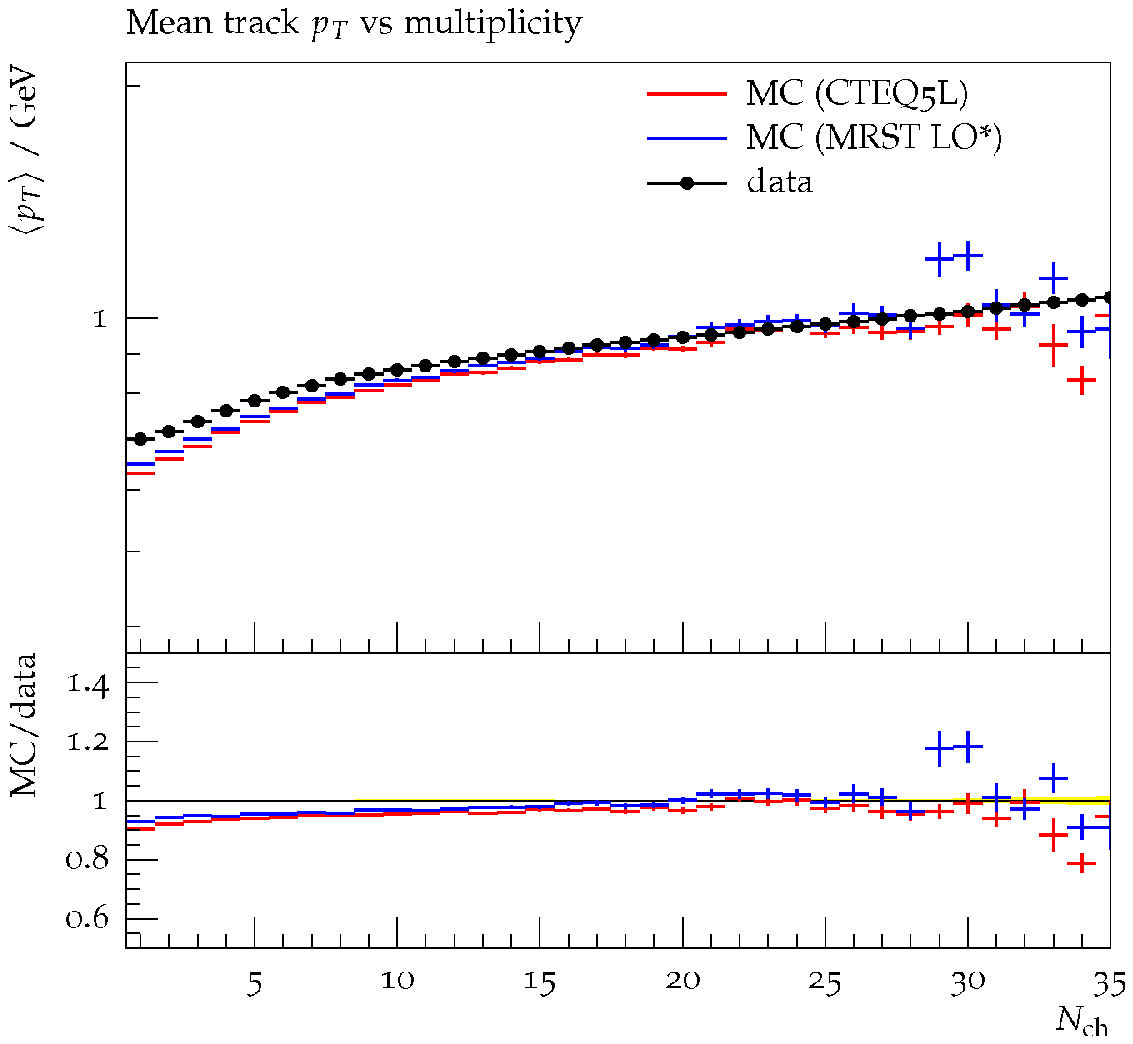}} 
  \subfloat[]{\label{fig:pTCMult-Riv2}\includegraphics[width=0.5\textwidth]
    {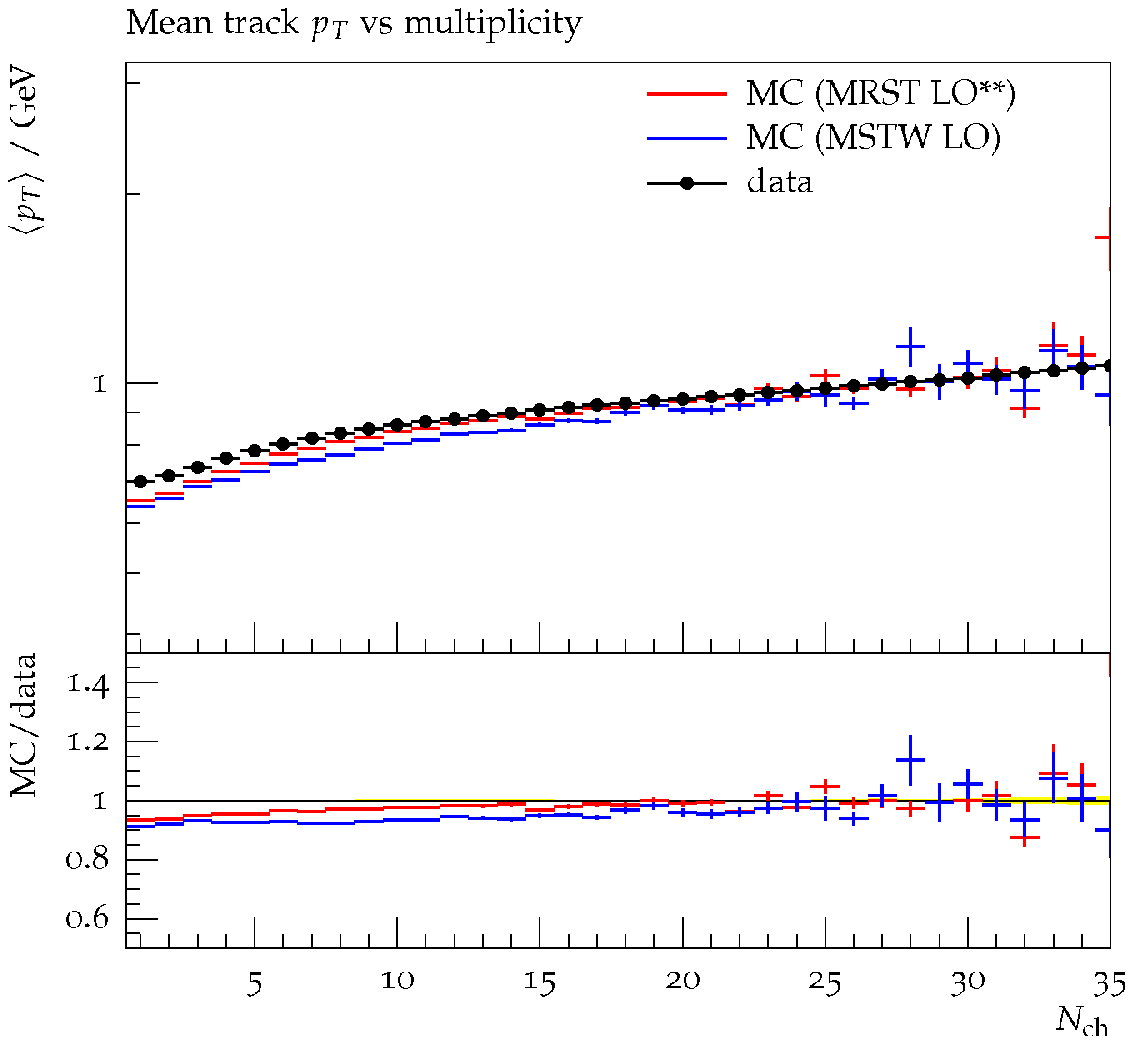}} 
  \caption{Evolution of the average transverse momentum, $<p_T>$, with charge
    multiplicity, $N_{ch}$, from the CDF Run 2 experiment compared with
    simulations with different PDFs}
  \label{fig:pTCMult-Riv}
\end{figure*}

Although different PDFs result in differences in many observables, they sometimes give more similar results. For example the evolution of the average transverse momentum with charge
multiplicity reproduce the data fairly well, independent of the choice of PDF. They all give slightly too low $\langle p_{\perp}\rangle$ at low
multiplicity and then increase relative to data so that they get closer as the
multiplicity increases, see Fig.~\ref{fig:pTCMult-Riv}. The only PDF that gives a
slightly different evolution is MSTW LO.

\section{Inclusive Jet Cross Section}
\label{sec:jet}
We compared the inclusive jet cross sections from our MC simulations with data 
collected by the CDF experiment at Tevatron Run 2 \cite{Aaltonen}, over five
pseudorapidity intervals ranging up to $\eta \leq 2.1$. In the experimental
analysis the jets are identified with the midpoint cone algorithm and also
compared to results with the $k_T$ algorithm  \cite{Salam}. We also examined the rapidity, multiplicity and transverse momentum
distributions for the individual hadrons. Two rapidity intervals and the transverse momentum distribution have been selected to show the cross section and illustrate our findings.

\begin{figure*}[tp]
  \centering
  \subfloat[]{\label{fig:jetCross2b}\includegraphics[width=0.5\textwidth]
    {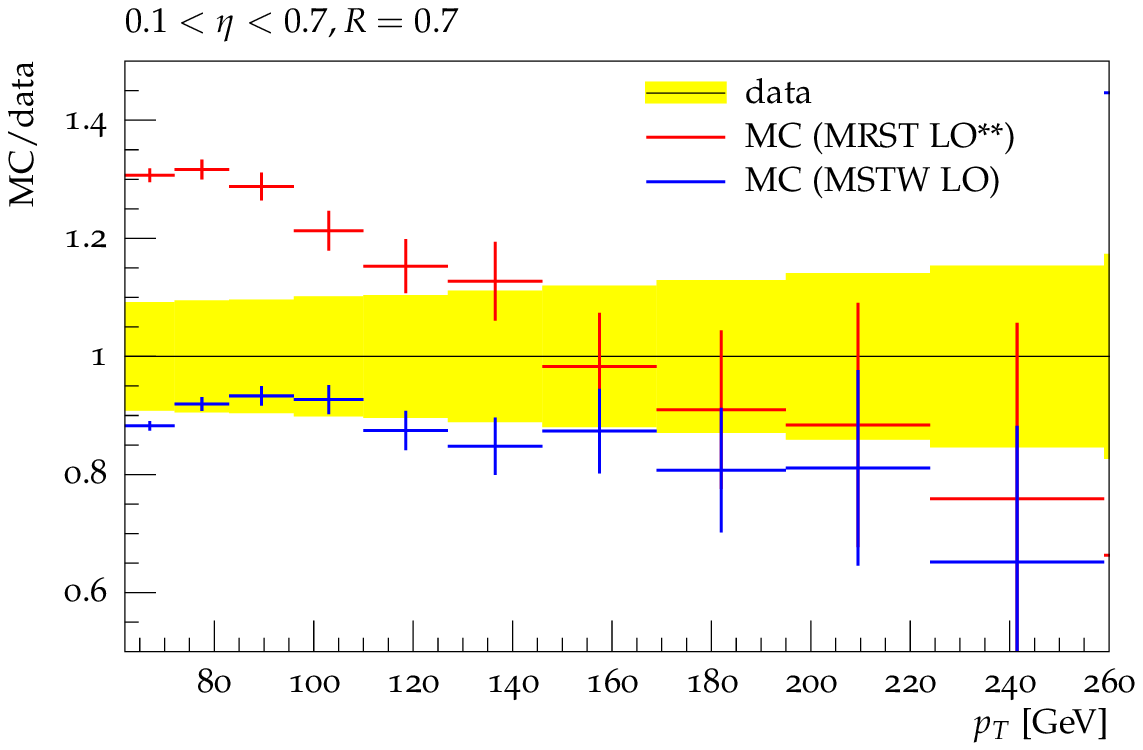}} 
  \subfloat[]{\label{fig:jetCross2d}\includegraphics[width=0.5\textwidth]
    {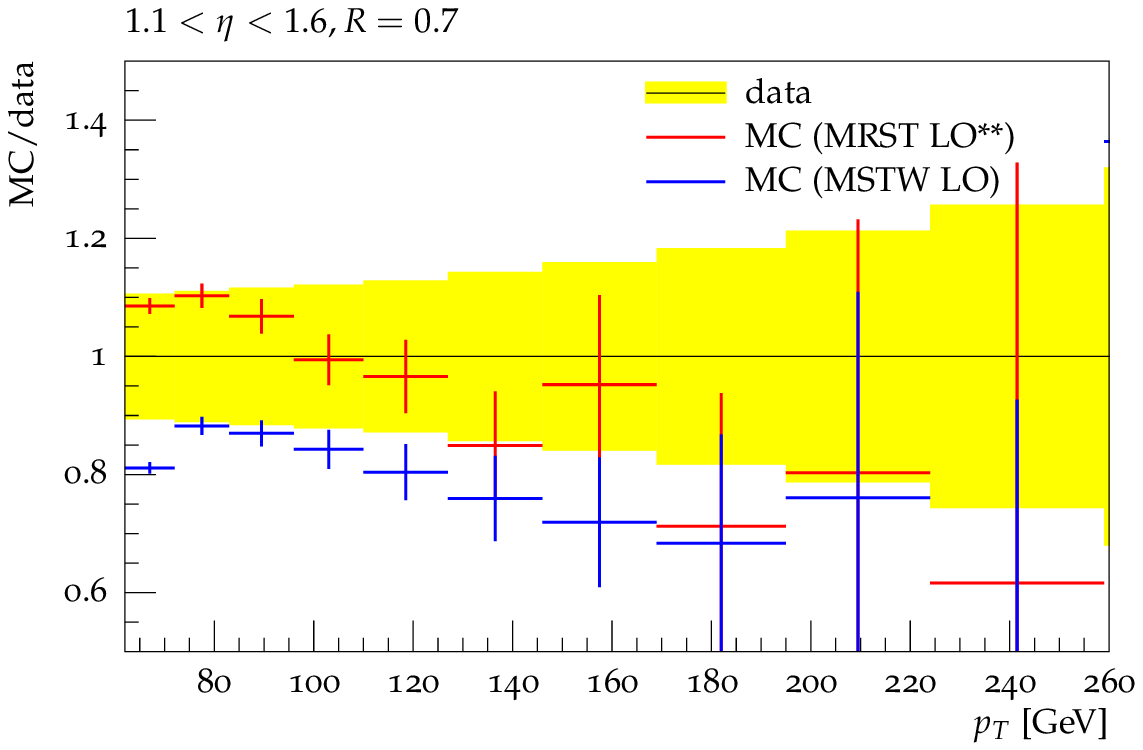}} \\
  \subfloat[]{\label{fig:jetCross4b}\includegraphics[width=0.5\textwidth]
    {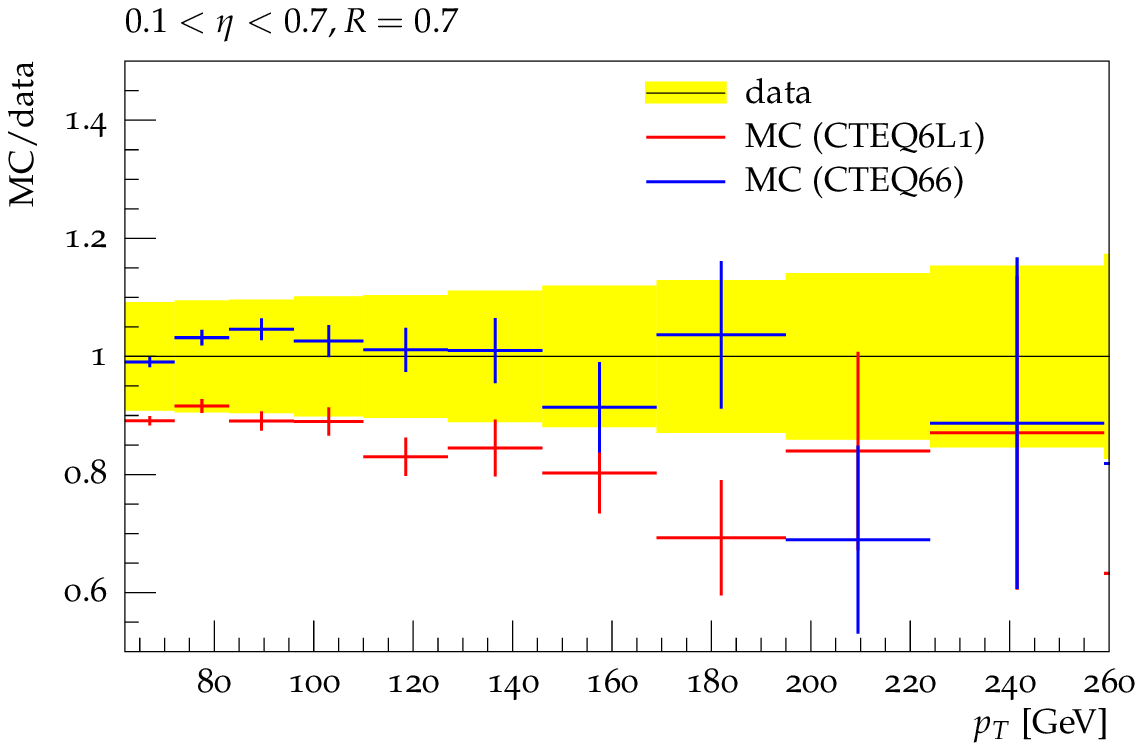}} 
  \subfloat[]{\label{fig:jetCross4d}\includegraphics[width=0.5\textwidth]
    {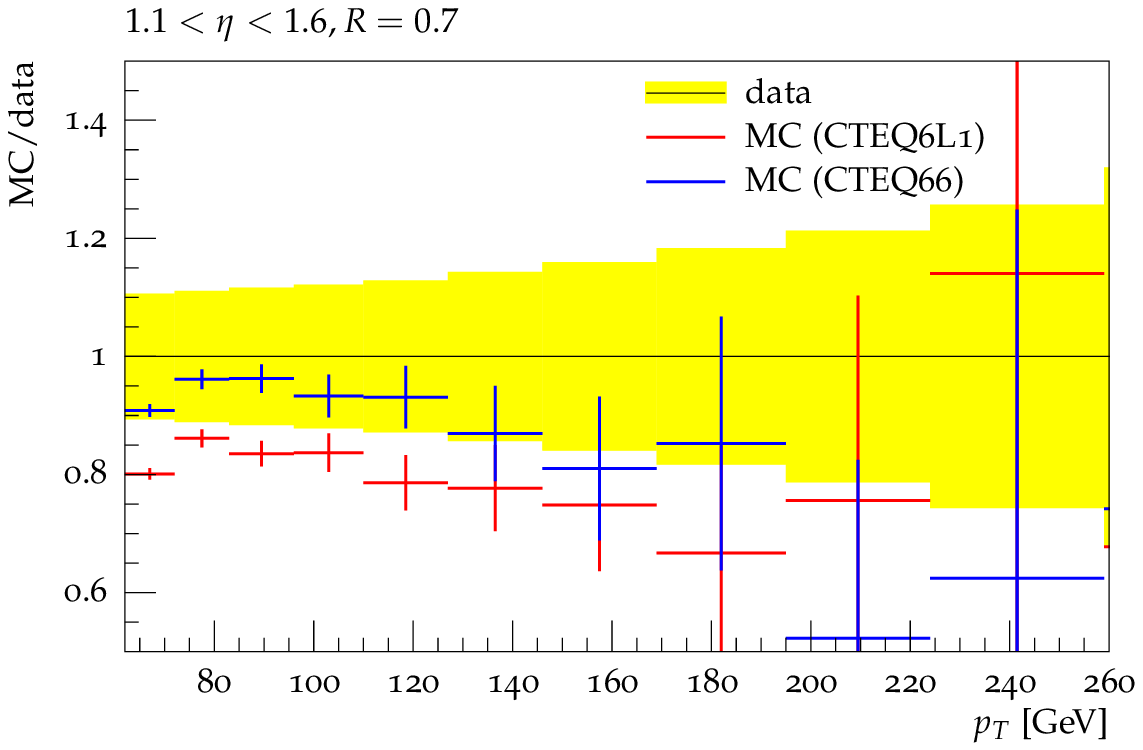}} \\
  \subfloat[]{\label{fig:jetCross5b}\includegraphics[width=0.5\textwidth]
    {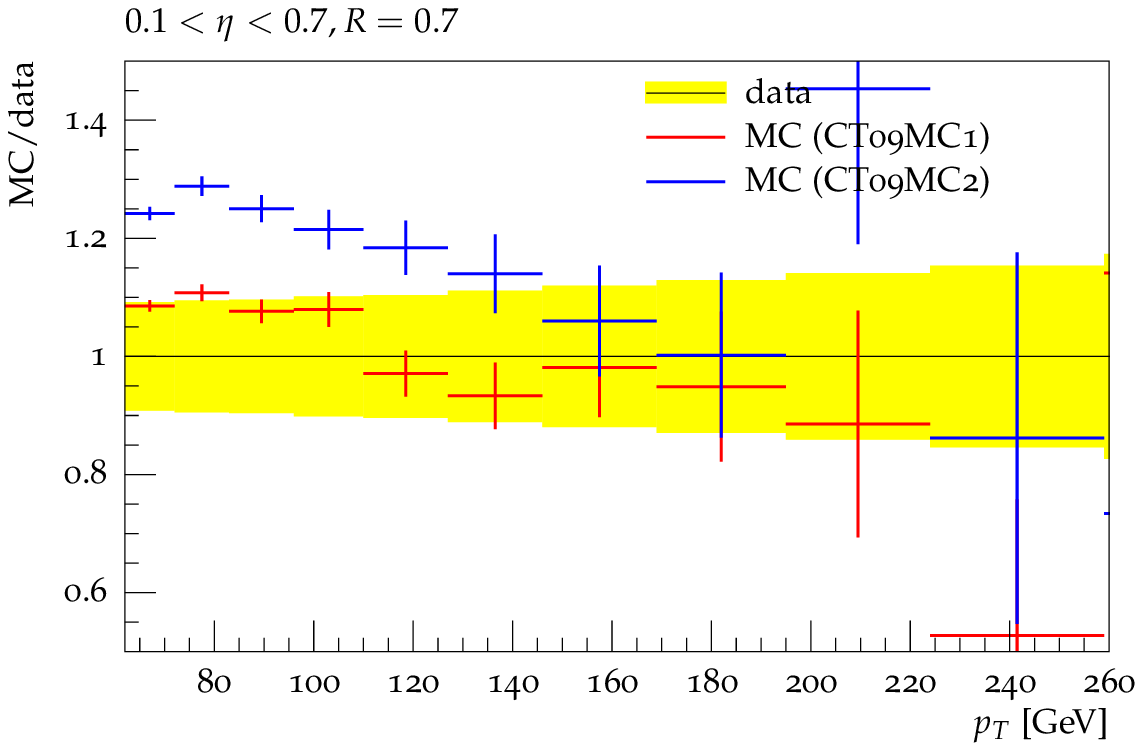}} 
  \subfloat[]{\label{fig:jetCross5d}\includegraphics[width=0.5\textwidth]
    {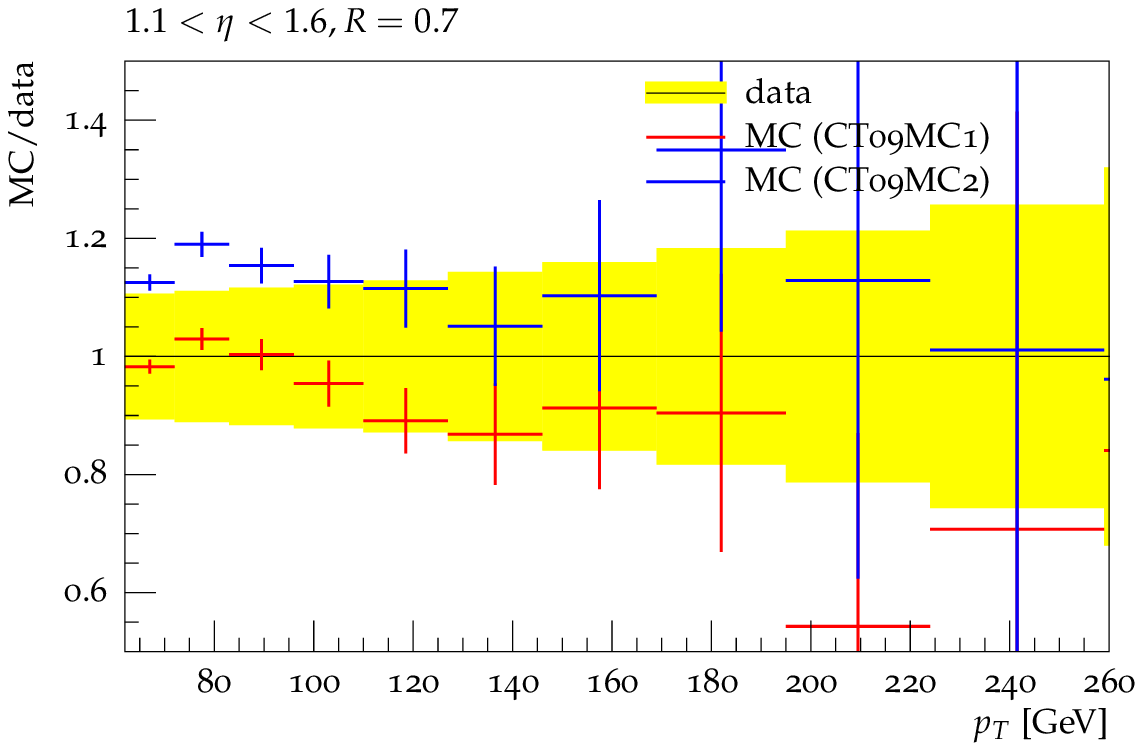}} 
  \caption{Ratio of the inclusive jet cross section, MC/Data, vs transverse momentum for two
    representative pseudorapidity intervals}
  \label{fig:jetCross2}
\end{figure*}

The inclusive jet cross section drops rapidly with increasing $p_{\perp}$ and spans over several orders of
magnitude. Since this makes differences between experiments and
simulations difficult to distinguish, we only show the
MC/data ratio in the following figures. The results with MRST LO**, MSTW LO,
CTEQ6L1, CTEQ66, CT09MC1 and CT09MC2 are shown in
Fig.~\ref{fig:jetCross2}. 

Generally the LO PDFs give similar results, as do the MC-adapted PDFs and the NLO PDFs are similar as well. MRST LO** starts with a much too large cross section and the
ratio quickly decreases when $p_{\perp}$ rises. This behavior is the strongest for LO** at low pseudorapidity; at larger $\eta$ the ratio gets smaller and flatter. All MC-adapted PDFs,
except MCS, show this type of behavior. MC2 and MC1 give results with very
similar shapes but the
MC2 cross section is larger, and MRST LO** and LO* are related much in the same fashion. MSTW LO and CTEQ6L give
cross sections which have similar behavior, and the ratio is much less dependent on
$p_{\perp}$ than with the
MC-adapted PDFs. CTEQ6L1 gives a ratio
which starts to decrease with $p_{\perp}$ at larger rapidities. CT09MCS gives a too low cross section, is once again different from the other MC-adapted PDFs, and gives results
which behave in a way more similar to those of the normal leading-order
distributions.  In the central rapidity regions the NLO PDFs actually give the cross sections closest to data with a ratio close to $1$, but their ratios decrease towards $0.5$ at larger
$\eta$. 

\section{Summary and Outlook}
\label{sec:sum}
Including the very latest parton distribution functions into \textsc{Pythia8}
both caused some technical troubles and gave some surprises,
especially while venturing outside the grid of the PDFs. At several occasions we were reminded that it
can be risky to use NLO PDFs in LO MC generators. In addition we found that there is a need for improved numerical stability at large
$x$, in order to keep the leading-order PDFs from going
negative. 

 There is also a need for better understanding of parton distribution functions at
small $x$, where the PDFs are 
now very different from each other. Excluding the NLO PDFs, the MRST/MSTW distributions are much larger than the
ones from CTEQ, and MSTW LO goes sky high compared to all other PDFs. At larger $Q^2$
the differences are smaller between the two collaborations, except for the
distributions that freeze their values at the end of the grid, and for MSTW LO
which is
still much larger than the rest. The quarks, and in particular
the up distributions, at large $x$ show smaller differences, but MSTW LO is once
again larger at small $x$.

The large differences in the PDFs get blurred when looking at
simulation results, but do nonetheless
 sometimes cause large
variations. The
MC-adapted PDFs frequently give results closer to data than the LO and NLO PDFs in their
respective group. The CTEQ PDFs have distributions and give results that are more
homogeneous, while
the MRST/MSTW PDFs give a broader spectrum. The MC-adapted PDFs from MRST give
results that show resemblance to results with the CTEQ distributions.

MSTW LO have a much larger gluon distribution at small $x$, which affect
results of, especially, minbias simulations. The rapidity distribution is
wide at the parton level ($2\rightarrow 2$ subprocess) and at hadron level lack the inward dents at larger rapidities, which all other PDFs give. The $p_{\perp}$ and $\sum E_{\perp}$
spectrum with this PDF have large deviations from experimental data. The
inclusive jet cross section is too low and the ratio to data is fairly constant with both
$p_{\perp}$ and $\eta$. The multiplicity increases with energy at a
more rapid rate than for all other PDFs. 

CTEQ6L and CTEQ6L1 have a larger gluon distribution than the CTEQ 
MC-adapted
PDFs at small $x$ for low $Q^2$, down to the freezing point,
but much smaller than MSTW LO. CTEQ6L cause a rapidity distribution with a narrow peak at parton level and a slightly larger peak value at hadron
level. The two distributions give similar results, but 6L is usually closer
to data and hence also closer to results with the MC-adapted PDFs from
CTEQ. 

The two NLO PDFs, MSTW NLO and CTEQ66 are similar and also give similar
results to each other for almost all observables in this study. They have the small gluon
distributions characteristic of NLO PDFs, and give rapidity distributions with a
low peak value at central rapidities. They are
generally further away from the data in the minbias simulations, and give a slower
multiplicity evolution with energy than the LO PDFs. Their
inclusive jet cross section is close to data at small pseudorapidities but too
low at large.

The MC-adapted PDFs which break the momentum sum rule, i.e. MRST LO*/**
and CT09 MC1/2, generally give
similar results to each other. They give high and narrow peaks in the rapidity
distributions and are generally closer to data than the ordinary LO and NLO PDFs from
their respective collaboration. The main exception is the inclusive jet
cross section at low pseudorapidities and $p_{\perp}$, where they give too
large values. Here they also give a relative decrease with $p_{\perp}$, a
rather surprising trend which is most
prominent for LO* and LO**, especially at low pseudorapidities. In the simulations of hard QCD events rapidity distributions
with the MC-adapted PDFs were narrower, and their multiplicity
distributions were shifted to lower multiplicity. Interesting to note is that the CT09MCS seems to
have some of the features of the other MC-adapted PDFs, but in some contexts
gives results more similar to ordinary leading-order PDFs.

For the leading-order PDFs a constant $K$-factor could improve the fit 
to the inclusive jet data, but for the
MC-adapted PDFs the difference in shape makes it more complicated. Differences in the PDFs have a larger impact when the CM-energy of the
collisions increases, and this can cause large uncertainties in
simulations at LHC energies. 

Many of the differences found can be explained by the differences in the gluon distribution at small $x$, where we see that the middle way represented by the MC-adapted PDFs give results closer to data for many observables. It is reasonable to suspect that the results with the LO PDFs from CTEQ would resemble those of MSTW LO if extrapolated towards small $x$.

 At this point there is no final answer as to which PDF gives the best results. In order to answer this question a much broader spectrum of observables is needed and complete
tunes for the different PDFs. This study does however highlight some of the relative differences between the PDFs when they are used under comparable conditions.

During the last years there has been a renewed interest in LO tunes with focus on
the applicability in MC generators. The MC-adapted PDFs resulted in some remarkable differences compared with leading-order PDF. They give results that are closer to data for many observables, although not for all. Thus there is space for further improvements. With these new PDFs 
a broader spectrum of tools is gained in \textsc{Pythia8} and in examining the origin of
differences and similarities between simulations and
experiments. Investigations are likely to continue in relation to LHC physics.


\end{document}